\documentclass[11pt,a4paper]{article}
\usepackage{graphicx}
\usepackage[latin1]{inputenc}
\usepackage{framed}
\usepackage[english]{babel}
\usepackage{amsmath}	
\usepackage{amssymb}	
\usepackage{bbm}
\usepackage{hyperref}
\usepackage{pb-diagram}

\def\be{\begin{equation}}
\def\ee{\end{equation}}
\def\ba{\begin{eqnarray}}
\def\ea{\end{eqnarray}}

\newcommand{\ket}[1]{| \, #1\rangle}

\begin{document}

\vspace*{-1cm}
\thispagestyle{empty}
\begin{flushright}
LPTENS 09/34 \\
RUNHETC-2009-28
\end{flushright}
\vspace*{1.5cm}

\begin{center}
{\Large 
{\bf Defect loops in gauged Wess-Zumino-Witten models}}
\vspace{2.0cm}

{\large C.~Bachas${}^{\, \sharp}$}%
\hspace*{0.3cm} {\large and} \hspace*{0.2cm} {\large S.~Monnier${}^{\,\natural, \, \flat}$}%
\vspace*{0.5cm}

${}^\sharp$ Laboratoire de Physique Th\'eorique de l'Ecole 
Normale Sup\'erieure 
\footnote{Unit\'e mixte de recherche (UMR 8549)
du CNRS  et de l'ENS, associ\'ee \`a l'Universit\'e  Pierre et Marie Curie et aux 
f\'ed\'erations de recherche 
FR684  et FR2687.}
\\
 24 rue Lhomond, 75231 Paris cedex, France\\
 bachas@lpt.ens.fr\\
\vspace*{0.4cm}

${}^\natural$ New High Energy Theory Center, Rutgers University \\
126 Frelinghuysen Road, Piscataway, NJ 08854, USA \\
monnier@physics.rutgers.edu \\ 

\vspace*{0.4cm}

${}^\flat$ On leave from Universit\'e de Gen\`eve, Section de Math\'ematiques\\
2-4 rue du Li\`evre, Gen\`eve 24,  Switzerland\\

\vspace*{2cm}

{\bf Abstract}
\end{center}

We consider loop observables in gauged Wess-Zumino-Witten models, and study the action of renormalization group flows 
on them. In the WZW model based on a compact Lie group $G$, we analyze at the classical level how the space of renormalizable
 defects is reduced upon the imposition of global and 
 affine 
 symmetries. We identify families of loop observables which are
   invariant with respect to  
   an affine symmetry  corresponding to a 
   subgroup 
  $H$ of $G$,  and
  show that they descend to gauge-invariant defects in 
     the gauged model based on $G/H$. We study the flows acting on
   these families perturbatively, and quantize the fixed points of the flows exactly. From their action on boundary states, we present a 
   derivation of the ``generalized Affleck-Ludwig rule'', which describes a large class of boundary renormalization group flows in rational conformal field theories. 

\newpage

\tableofcontents

\newpage 


\section{Introduction and summary of results}

Two-dimensional field theories have a rich set of interesting loop observables,  much richer than  in higher
dimensions. The
loop operators of $2D$  conformal field theories (CFT's), in particular, whose study was pioneered by Bazhanov,  Lukyanov and Zamolodchikov  
 \cite{Bazhanov:1994ft, Bazhanov:1996dr, Bazhanov:1998dq},
 have  attracted  considerable  attention. Some of these operators  
 describe interesting condensed-matter systems \cite{Kane:1992zza, Oshikawa:1996dj, Fendley:2009gm}.  
 They have  furthermore proved to be powerful tools for organizing  boundary renormalization-group flows \cite{Graham:2003nc, Bachas:2004sy, Fredenhagen:2009tn, 
  Monnier:2009ag, Brunner:2009zt, Kormos:2009sk},   
and they could play  a role as symmetries of  string theory \cite{Bachas:2008jd, Frohlich:2004ef, Frohlich:2006ch, Brunner:2008fa}. 
A partial  list of further references is \cite{Petkova:2000ip, Quella:2002ct, Quella:2006de, Alekseev:2007in,  Mikhailov:2007mr, Brunner:2007qu, 
 Runkel:2007wd, Bachas:2007td, Brunner:2007ur, Gang:2008sz, Sarkissian:2008dq,  Sarkissian:2009aa}. 
 Our aim in the present work  will be  the  study of  loop operators in the largest known class of exactly-solvable
conformal theories,  which includes  all known rational CFTs:    
  the Goddard-Kent-Olive (GKO) coset models \cite{Goddard:1984vk, Goddard:1986ee}. 

 \vskip 0.5mm

The loops under study arise as worldlines of point-like defects,  or  ``quantum impurities''. 
An impurity is  characterized by a state space $V$, that we will choose to be finite-dimensional,  and by a Hamiltonian ${\cal H}_{\rm imp}$ 
which is a ${\rm dim}(V)\times {\rm dim}(V)$ matrix. 
This latter depends  in general on the local bulk fields, which we denote collectively by $\Phi$.  
 The classical  loop  observables are given by 
\be
\label{FormLoopOp}
{\cal O}(C)  =   {\rm tr }_V  \,  P \, e^{i\oint_C  ds \,  {\cal H}_{\rm imp}(\Phi) }\ ,
\ee
where $s$ parametrizes the loop $C$,    and $P$ stands for path ordering.  
In what follows we will mostly work on the cylinder  $\mathbb{R}\times S^1$, and let $C$  
wind once around the cylinder. 
Strictly-speaking,   the worldline of an impurity must be time-like, in which case it cannot possibly be a loop.   
 We may however interpret ${\cal O}(S^1)$ in the Euclidean theory  as   the insertion of a (probe)  defect  at finite
temperature. Alternatively, one may take $\mathbb{R}$ to be the time direction, and think of  ${\cal O}(S^1)$  
as a fixed-time observable.  Both of these interpretations are familiar from the study of the Wilson and  Polyakov loops in
ordinary  gauge  theories. 
 
 \vskip 0.2mm
 
Expression (\ref{FormLoopOp})  does not,  in general,  make sense  after  the bulk fields $\Phi$ have been quantized, 
 because  of short-distance divergences.  The operators  ${\cal O}(C)$ must  be first regularized and then renormalized,  and  this
 induces a flow in the space of  impurity Hamiltonians.  Computing these renormalization-group  
 flows is one of the  central problems in the subject.
The fixed-point operators can  
often be  found exactly  by algebraic methods (see e.g.  \cite{Petkova:2000ip, Quella:2002ct, Alekseev:2007in,  Monnier:2009ag}), 
while in special situations 
(and for specific renormalization schemes) the full flow may be  
  integrable.\footnote{ This is the case
for the   minimal-model  loop operators of Bazhanov et al  \cite{Bazhanov:1994ft, Bazhanov:1996dr, Bazhanov:1998dq}, whose
structure was  further elucidated by Runkel \cite{Runkel:2007wd}. The early motivation for this work was to explore the
integrable structure of the bulk CFT.  The connection between loop operators and bulk integrability is easy to understand at the
classical level:    the defect Hamiltonians of interest  are connection forms  on a bundle over two-dimensional spacetime with fiber $V$.   When  the  equations  of motion  imply the
flatness of the connection,  ${\cal O}(C)$  depends only on the homotopy class   of $C$ and is therefore
 an integral of motion.  Integrable field theories usually possess a  continuous family of such defects, 
 parametrized by an arbitrary coupling $\lambda$,  and which  generate   an infinite number 
of  (not necessarily independent)  conserved charges. 
In the language of integrable systems  \cite{citeulike:2386224} one says that ${\cal H}_{\rm imp}$ is derived from a Lax connection, 
$V$ is  the auxiliary space,  $\lambda$   the spectral parameter, and ${\cal O}(C)$  the trace of the associated 
monodromy matrix.  After quantization, $\lambda$  generally runs -- for an interesting
exception see \cite{Mikhailov:2007mr}. 
} 
In  general, however, the only available analytical  tool is perturbation theory. A method to compute the RG flows for a class
of  loop operators in the weakly-coupled  Wess-Zumino-Witten (WZW) models has been proposed
in ref. \cite{Bachas:2004sy}.  The key  idea is to construct  the regularized and renormalized  operators  
as elements in the enveloping algebra of the
current algebra,   $U (\hat{\mathfrak{g}}_k)$,  using an  expansion in inverse powers of  the level $k$.  In this paper
 we will show how to extend this  approach to the  weakly-coupled  GKO  models. 

 \vskip 0.8mm

 The most  general,  renormalizable by  power-counting and  classically scale-invariant defects  in  two-dimensional  $\sigma$-models
 depend on a number of arbitrary functions on the target manifold. It is possible to reduce this infinite parameter space by imposing extra symmetries. 
  For a WZW model with Lie group $G$,    the symmetry of the defect  must be  a subgroup 
 of the affine $\widehat G_{\rm left} \times \widehat G_{\rm right}$  bulk symmetry.  Our first task, in section
  \ref{SecDefLoop}  below, will be
 to analyze the various   possible reductions. As we will see,  
   global symmetries that act transitively on the target space, 
 such as $G_{\rm left}$,  suffice  to restrict  ${\cal H}_{\rm imp}$ to a finite-dimensional parameter space.   
However, despite this huge reduction,   the   renormalization of  ${\cal O}(C)$   remains   in general an 
arduous task.  
A further simplification occurs when we extend the invariance under $G_{\rm left}$ to the full loop group $\widehat G_{\rm left}$ and assume that the latter acts trivially on $V$. Defects satisfying these two properties couple 
 only to  the right-moving sector of the theory, and will be referred to as 
 \emph{holomorphic}.\footnote{Holomorphic loop operators form a special subclass of the \emph{chiral} loop operators, defined in  \cite{Bachas:2004sy} as  the loop operators commuting with the left   Virasoro algebra. 
  } 
  Their Hamiltonian is parametrized by
   ${\rm dim}(\mathfrak{g})$ constant hermitean matrices, $M^a$,  and 
  depends only on the right-moving WZW currents ${\cal J}_-^a$,
    \be
   {\cal H}_{\rm imp} =  - {1\over k} \sum_a  M^a {\cal J}_-^a \ . 
   \ee
We   henceforth  focus our attention on holomorphic  defects, which
can be quantized algebraically, along the lines of  ref. \cite{Bachas:2004sy}. This is  a restriction of
 convenience, not of   higher principle. 

 \vskip 1.5mm

 The  parameter space  of  holomorphic defects  can be  further reduced by
 imposing invariance under a global subgroup 
 $H_{\rm right} \subseteq G_{\rm right}$ or a loop subgroup $\widehat H_{\rm right} \subseteq \widehat G_{\rm right}$. 
However, while the global  symmetry is manifest,  the  affine symmetry associated with $\widehat H_{\rm right}$  is a priori  broken by the  
  ultraviolet cutoff.  We will nevertheless   argue (but  will not prove)  that a  $\widehat H_{\rm right}$-invariant subspace of parameter space is present at the quantum level. 
This will bring us to the main claim  of this  paper:  \vskip 1.9mm

\noindent {\it  The  holomorphic,  $\widehat H_{\rm right}$-invariant   defects of the {\rm WZW}  model with group $G\supseteq H$ can be mapped to  \emph{local}  defects, with  \emph{identical} RG  flows,  in the  gauged {\rm WZW} model based on $G/H$.} 
   
   \vskip 1.9mm

\noindent Notice that in the GKO construction of the state space of the coset model, one also starts from the state space of the parent  WZW model, which is then  
projected to ($\widehat H_{\rm left}\times \widehat H_{\rm right}$) - invariant states  \cite{Goddard:1984vk, Goddard:1986ee, Bowcock:1988xr}. What we will show in this paper
 is that a similar  procedure works for (a class of)  impurities and for their RG flows. 
The result  is not trivial  because the gauged and ungauged   WZW models are related by a  
non-local   transformation of fields \cite{Gawedzki:1988hq, Gawedzki:1988nj, Karabali:1988au}.   
 
 \vskip 0.5mm

Let us now describe in more detail this reduction. 
As will be derived in section \ref{SecRenBeta}, the RG flow of holomorphic defects in the WZW model is given to leading order in $1/k$ by
 \be
\label{BetaFun}
  \frac{d M^a}{d\,  {\rm log}\epsilon} =    \sum_{b,c} \frac{1}{2k} \left [ M^b,\   i f^{abc} M^c  -   [M^a, M^b] \right ] \; +\;  O({1/ k^2})\,  , 
\ee
where $\epsilon$ is a short-distance cutoff  
and $f^{abc}$ are the structure constants of the group $G$. 
These first-order  equations describe a  {\it gradient}  flow, i.e.  their 
 right-hand-side is the variation of an action (the effective open-string action of \cite{Alekseev:2000fd}): 
\be
\label{EffOAct}
S(M) \, =\,   - \frac{1}{8k} \sum_{a,b} \mbox{Tr}([M^a,M^b]^2) +  \frac{1}{6k} \sum_{a,b,c} if^{abc} \mbox{Tr}(M^a[M^b,M^c])\; +\;  O({1/ k^2})\, \;  .  
\ee
 This action was  studied extensively by one of us \cite{Monnier:2005jt}.  Its reduction to
 defects with a global  symmetry $H\subseteq G$ proceeds in two steps: first specify 
 how  $H$  acts on $V$, i.e. choose a 
 representation $R$ of $H$ with ${\rm dim}(R) = {\rm dim}(V)$, and then require the 
matrix-valued vector $M$ to be an invariant $H$-tensor.\footnote{Except in  section \ref{SecDefLoop},
 the defects considered in this work  couple only to the right-moving sector. To lighten the
 notation,  and when no confusion is possible, 
 we  drop the subscript ``right" from the symmetry groups.} 
The number of free parameters, for a given $H$ and $R$,  
is  equal to  the number of  trivial representations  
 in the decomposition of $R\otimes R^*  \otimes \mathfrak{g}$. (Here  $\mathfrak{g}$ is
 considered as a representation of $H$, and $R^*$ is the conjugate of $R$.)
 \vskip 0.5mm

We will decompose the Lie algebra as $\mathfrak{g}= \mathfrak{h} \oplus \mathfrak{g}/\mathfrak{h}$ and use an orthonormal basis $\{e^a\}$ of $\mathfrak{g}$ compatible with this decomposition. Indices $i$,$j$,... will run over the generators of $\mathfrak{h}$, while indices $s$,$t$,... will run over generators of $\mathfrak{g}/\mathfrak{h}$. It is useful to give separate names to the $H$-invariant couplings of the corresponding currents,
   \be \label{globalRed}
   {\cal H}_{\rm imp} =  - {1\over k} \sum_{a}   M^a {\cal J}_-^a  =   
    - {1\over k} \left(  \sum_{j} \Theta^j {\cal J}_-^j + 
     \sum_{s} \tilde  \Theta^s {\cal J}_-^s \right) 
    \ , 
   \ee
   where $\Theta$ is an invariant tensor in $R \otimes  R^* \otimes \mathfrak{h}$ and $\tilde  \Theta$ an invariant tensor in $R \otimes  R^* \otimes \mathfrak{g}/\mathfrak{h}$.
 If  $H=G$ and $R$ is  irreducible, 
the invariant Hamiltonian is given by  $M^a =  \lambda\,  T^a$  with $\lambda$ real and $T^a$ the 
 generators of $G$  in the representation $R$. This is the one-parameter reduction
 of the RG  flow   analyzed  in ref. \cite{Bachas:2004sy}. 
 \vskip 0.7mm

 The above parameter space of $H$-invariant holomorphic defects 
 is further reduced if one imposes invariance under the action of the loop group $\widehat H$. There is a distinguished
  invariant tensor $\Theta_R$ in $R \otimes R^\ast \otimes \mathfrak{h}$, namely the one whose matrix elements coincide with those of the generators of $H$ in the representation $R$, normalized with respect to the Killing form of $\mathfrak{g}$.  
   We will show that in the classical theory, the condition of $\widehat H$ invariance reads:
  \be\label{red6}
    \Theta = \Theta_R \;. 
 \ee
However, while   \eqref{globalRed} is a consistent truncation of the complete RG  flow equations,
the reduction  \eqref{red6} is in general inconsistent at higher orders
in the $1/k$ expansion.  
 We will argue that the 
affine  symmetry is, nevertheless,  preserved   on a subspace ${\Sigma}(\widehat H,R,k)$ of the space of invariant tensors, which is a (small at large $k$) deformation of the classical subspace  \eqref{red6}.
Its precise form  depends   on  the renormalization scheme. 
 Proving the existence of this invariant subspace
 at all orders in the $1/k$ expansion  is an interesting open mathematical problem. 
 Assuming that it exists,   we can identify the RG flows  in ${\Sigma}(\widehat H,R,k)$
with the  flows  of holomorphic defects in the $G/H$ coset model. 

  \vskip 1 mm

  The RG flow of a holomorphic 
  defect  can be imprinted on other defects, or on boundaries,  
through {\it fusion}.  This has proved to be  a convenient way of organizing boundary RG flows  \cite{Graham:2003nc, Bachas:2004sy}.
 Fusion is the operation that merges the {\it parallel} worldlines of two defects, so that the region
in between them shrinks to zero.
 Similarly, a defect loop can be fused with a parallel boundary. 
 These operations  are in general singular  (see ref. \cite{Bachas:2008jd})
but  for the holomorphic (though not necessarily conformal) defects studied here the fusion is  smooth.\footnote{The key property
of the corresponding loop operators, which guarantees smoothness of their fusion,  is that they commute with rigid spacetime translations. 
In the special case of  topological defects, see section \ref{cctdefects}, the fusion is smooth for any pair of homotopically-equivalent
worldlines, even if they are not parallel.}
  The RG flows  between holomorphic defects given by  \eqref{BetaFun}  
 can therefore be imprinted smoothly on boundaries or on other defects  --  they are in this sense  {\it universal} 
 flows.

  \vskip 0.5 mm 
  
  A  simple illustration is provided by the boundary Kondo flows in WZW models.  These describe the
  screening of a boundary ``spin", $S$, whose coupling to the bulk currents is  ${\cal H}_{\rm bnry} = \lambda S^a J^a$. 
    Affleck and Ludwig formulated a general
   rule   \cite{Affleck:1990by}  to determine the  IR fixed point of the  flow,  
    in terms of the UV fixed point and  the boundary spin.  This so-called
       ``absorption-of-boundary-spin'' principle, stated originally  for the physically most relevant case $G=SU(2)$,
 can be written succintly as follows: 
\be
\label{ALRule}
{\rm dim}(S) B_\mu \rightarrow \sum_\nu \mathcal{N}_{\mu\sigma}^{\;\;\nu} B_\nu \ .
\ee
Here $\mathcal{N}_{\mu\sigma}^{\;\;\nu}$ are the fusion coefficients of the WZW model, 
$\sigma$ is the highest weight of the representation  $S$ of ${\mathfrak{g}}$, and
   the $B_\mu$ are maximally symmetric  boundaries (on which left and right currents are identified) 
   labeled by  integrable highest weights
    of the Kac-Moody algebra $\hat{\mathfrak{g}}_k$ \cite{Cardy:1989ir}.  
    It was shown in \cite{Bachas:2004sy} that the Kondo  flows 
 are all  imprints of the universal flows acting on holomorphic defects,  
 \be
\label{FlowOp}
{\rm dim}(S) \mathbbm{1} \rightarrow O_\sigma\ ,  
\ee
where the UV operator on the left  corresponds to an impurity with  dim($V$) = dim$(S)$ and ${\cal H}_{\rm imp} = 0$, 
 while the  IR operator
 on the right   commutes with the entire affine  algebra, and $O_\sigma$ is 
 the quantum version of the trace of the classical 
  monodromy.\footnote{For semiclassical derivations
 of the quantum monodromies  see  \cite{Gepner:1986wi,Alekseev:1990vr} and the references
  in section \ref{SecRevGaugWZW}.}
 This  IR operator  can be  
 constructed explicitly as an  element of a completion of the enveloping algebra $U(\hat{\mathfrak{g}}_k) $
   \cite{Alekseev:2007in} and it obeys:   
\be\label{fusion}
O_\sigma O_\mu = \sum_\nu {\cal N}_{\sigma\mu}^{\;\;\nu} O_\nu \ , \hskip 1cm {\rm and}\hskip 0.6cm
O_\nu B_0 = B_\nu\ . 
\ee
 \noindent  The Affleck-Ludwig rule follows easily from the above two relations.  
   \vskip 0.5 mm  
  
This rederivation of  the Kondo flows has some immediate advantages.  First, it shows that 
  the  $\beta$-function of the  flow  \eqref{ALRule}
does not depend on  the UV fixed point $B_\mu$.   Second, one can fuse the defect flow
 \eqref{FlowOp} with  other 
 (e.g. symmetry-breaking)   boundary states  to find new  boundary RG flows and fixed points \cite{Monnier:2009ag}.  In particular  some, 
  but not all,  of the RG  flows between twisted WZW boundary states \cite{Alekseev:2002rj}    
 can be obtained in this way. Finally, 
  one can derive relations  between different partition functions on the annulus by 
  freely  transporting a holomorphic defect from one of the boundaries to the other.     
   \vskip 0.5 mm

  The Affleck-Ludwig rule has been  generalized to  GKO coset models by Fredenhagen and Schomerus 
  \cite{Fredenhagen:2001kw, Fredenhagen:2002qn, Fredenhagen:2003xf}. Conformal boundaries of
 the  $G/H$ model are labelled by pairs $[\mu,\gamma]$ of integrable weights of the Kac-Moody algebras $\hat{\mathfrak{g}}_k$ and $\hat{\mathfrak{h}}_{xk}$,  
  modulo some selection and identifications (see \cite{Fredenhagen:2003xf} for details).  Fredenhagen and Schomerus proposed the 
  following set of boundary RG  flows:
\be
\label{5}
\sum_{\alpha,\gamma}b_{\sigma \alpha}\, \mathcal{N}_{\alpha\beta}^{(\mathfrak{h}) \;\gamma}\  B_{[\mu,\gamma]}
\  \rightarrow \  \sum_\nu \mathcal{N}_{\sigma\mu}^{(\mathfrak{g})\;\nu}\  B_{[\nu,\beta]} \ , 
\ee
where  $\sigma$, $\mu$, $\nu$ are weights of $\hat{\mathfrak{g}}_k$,   $\alpha$, $\beta$, $\gamma$ are weights of $\hat{\mathfrak{h}}_{xk}$, 
  $b_{\sigma \alpha}$ are the branching coefficients of the $\mathfrak{h}$-representation of highest weight $\alpha$
 in the $\mathfrak{g}$-representation of highest weight $\sigma$, and
  $\mathcal{N}^{(\mathfrak{h})}$ and  $\mathcal{N}^{(\mathfrak{g})}$ are the fusion rules of the corresponding affine
algebras.  
The reader can verify,  as  a check,   that   
\eqref{5}  reduces to (\ref{ALRule})  when $H$ is the
trivial subgroup of $G$.  
The above generalized  Affleck-Ludwig rule  reproduces a large class of known  
RG boundary  flows in minimal models and in parafermionic theories \cite{Fredenhagen:2003xf}.
    \vskip 0.5mm
 
 Part of our motivation for  the present work was  the wish to derive the flows  \eqref{5}  
 as imprints of universal defect flows,  by extending the corresponding analysis of the  Kondo problem.   
 We will argue that the flows  \eqref{BetaFun} restricted 
 to  the  $\widehat H$-invariant  defects  account for all the boundary flows 
 predicted by the generalized Affleck-Ludwig rule.  The existence of these RG flows, at  
 least  at this leading order in $1/k$,  can be established analytically.  Explicit solutions 
 of the coupled non-linear flow equations  can be, of course,  
 also obtained by numerical means. 
 
\vspace{1.5mm}

The rest of this paper provides the arguments and the detailed calculations supporting 
 these  claims. We begin, in section \ref{SecDefLoop}, with a general analysis of perturbatively-renormalizable defects in  the (ungauged)  WZW model.  We describe the reductions of parameter space when invariance under global or affine bulk symmetries are imposed on a defect.
 This section extends and rectifies a misleading point in the corresponding discussion of ref.  \cite{Bachas:2004sy}. 
 Section \ref{SecRevGaugWZW} starts with a brief review
 of classical gauged WZW models and their quantization. 
  We then go on to show how the holomorphic, $\widehat H$-invariant WZW defects are mapped
 to local, gauge-invariant defects of the coset model.  We also discuss the relation of
 special enhanced-symmetry defects  with classical monodromies. 
 In section \ref{SecPertQuant} we quantize the holomorphic defects in a perturbative expansion in $1/k$,   
using the algebraic method of ref. \cite{Bachas:2004sy}. We derive the flow equations \eqref{BetaFun}, 
analyze their fixed-point structure and solve them numerically in some simple examples. These examples allow
to visualize the invariant subspaces on which the generalized Kondo and the Fredenhagen-Schomerus 
flows are defined. Finally, in section \ref{SecNonPert} we first  use the enhanced symmetries of  (some of) the fixed-point
operators to calculate their exact quantum spectrum.   This is a straightforward  extension of  the results of \cite{Alekseev:2007in}. 
We then explain how the 
  Fredenhagen-Schomerus flows \eqref{5} can be obtained as
  imprints of  our universal defect-flow equations. 
   A technical point concerning the action of the fixed-point
  operators in the BRST quantization of the coset model   is treated 
  separately  in  appendix \ref{BRST} .

 
 \section{Symmetries of WZW defects}
\label{SecDefLoop}

 In this section we analyze the classical symmetries of the defect operators \eqref{FormLoopOp}. 
  We begin with a general discussion of defect loops and the conformal group, and then 
  proceed to examine the WZW defects and their possible global or affine symmetries. 
  Finally we  narrow down  to  the holomorphic defects,  which are the main focus in the rest of the paper.  
  This section  extends and clarifies in significant ways   the
   discussion of defect symmetries  of  ref.  \cite{Bachas:2004sy}.

\subsection{Conformal,  chiral and topological  defects}
\label{cctdefects}

The observables \eqref{FormLoopOp} are traced evolution operators for a quantum impurity moving along the trajectory $C$ and interacting 
locally with the fields in the bulk.  The latter  are for now  classical,  while the impurity is from the very start quantum.
We are interested in impurities which are 
\emph{scale-invariant}  at the classical level, so that ${\cal H}_{\rm imp}$ contains  no  dimensionful couplings.   
Renormalization may generate couplings with the dimension of mass. These are relevant in the infrared
and we will assume that they are tuned  to zero. 
 \vskip 0.5 mm   
   
   In a four-dimensional theory scale invariance is very restrictive: it forces ${\cal H}_{\rm imp}$ to be linear in the scalar and/or the gauge
    fields.\footnote{Scalar couplings enter for instance in the supersymmetric Wilson loop of $N=4$ super
 Yang-Mills \cite{Maldacena:1998im, Rey:1998ik}. We assume that the impurity has no internal bosonic degrees of freedom. 
 Fermionic degrees of freedom correspond to  a finite number of states which can be included in $V$.}  
In two dimensions, on the other hand, there is much greater freedom.  If the bulk theory is a non-linear (ungauged)
 $\sigma$-model with fields $\Phi$ parametrizing  a target manifold ${\cal M}$,   the most general classically scale-invariant defect Hamiltonian reads \cite{Bachas:2004sy}:
\be\label{gen1form}
  \int_C  ds\,   {\cal H}_{\rm imp} =   \int_C  d\zeta^\alpha \, [  \partial_\alpha \Phi\cdot {\mathbf{B}}(\Phi)   +
       \epsilon_{\alpha\beta} \partial^\beta  \Phi\cdot {\mathbf{C}} (\Phi)  ] \ . 
\ee 
Here $\zeta^\alpha$ are the coordinates  of the two-dimensional spacetime, and $\epsilon_{\alpha\beta}$ is the antisymmetric tensor. 
${\mathbf{B}} \cdot d\Phi$ and 
${\mathbf{C}}\cdot d\Phi$ are the pull-backs on the 2D spacetime of arbitrary matrix-valued one-forms ${\mathbf{B}}$ and ${\mathbf{C}}$ on the target ${\cal M}$.
There are no dimensionful parameters in 
${\cal H}_{\rm imp}$  because  $\Phi$ has dimension zero. 
It is convenient to introduce the short-hand notation
\be\label{Ww1}
 {\cal W}_\alpha \, d\zeta^\alpha
   :=  -  i ( \partial_\alpha \Phi\cdot {\mathbf{B}} + \epsilon_{\alpha\beta} \partial^\beta \Phi\cdot {\mathbf{C}} ) \, d\zeta^\alpha\ . 
\ee
We can  consider  ${\cal W}$ as a (composite)  matrix-valued connection form,  and the loop observables ${\cal O}(C)$ as the corresponding Wilson loops. However, no assumptions about the transformation properties of ${\cal W}$ are being made at this stage. Notice for later reference  that
in light-cone coordinates 
 the impurity ``Hamiltonian" is \hskip 2mm
${\cal H}_{\rm imp} =  i{\cal W}_+ \pm i{\cal W}_-$, according to whether $C$ is in the time
or in the space direction.

\vskip 0.6mm
 
 The scale invariance of  (\ref{gen1form}) extends to invariance under all conformal
  transformations which preserve the defect  worldline  $C$.  
 This symmetry is further enhanced if, as a result of the field equations, ${\cal W}$ turns out to define a flat connection, i.e. if
\be
\partial_\alpha \mathcal{W}_\beta  - \partial_\beta \mathcal{W}_\alpha + [ \mathcal{W}_\alpha , \mathcal{W}_\beta ]
= 0\ . 
\ee
In this case the non-abelian Stokes theorem implies  that ${\cal O}(C)$ is invariant under arbitrary continuous deformations of the curve $C$.
Such defects are therefore  {\it topological}, and on a cylindrical spacetime they
define a set of  dim$(V)$ conserved charges.
A continuous family of such defects gives an infinite number of integrals of motion and is  
usually tantamount to classical integrability (see for instance \cite{citeulike:2386224}). 
 
 \vskip 0.4mm 
Quantization breaks, in general, the scale invariance of the defect loop even when the bulk theory is conformal.   
This is because the definition of ${\cal O}(C)$ requires the introduction of a short-distance cutoff. 
As the cutoff $\epsilon$ is removed, the coupling functions $ {\mathbf{B}}(\Phi)$ and $ {\mathbf{C}}(\Phi)$   
flow either to infinity or to infrared fixed points where scale-invariance is restored.  The 
fixed-point operators  $O^*(C)$  commute with the  
diagonal subalgebra of the full  ${\rm Vir}_{\rm left} \oplus {\rm Vir}_{\rm right}$  conformal symmetry of the bulk.
More explicitly, if $\overline L_n$ and $L_n$ are the left- and  right-moving Virasoro generators on a cylindrical spacetime, then  
\begin{equation}
\label{Vir} 
[ \, L_n - \overline L_{-n}\,  ,  \,    O^*(S^1)\,  ] \, = \, 0 \ \ \ \ \ \mbox{for all} \; n\in {\mathbb Z}  \ .  
\end{equation}
Topological operators, first introduced in Conformal Field Theory  by  Petkova and Zuber \cite{Petkova:2000ip}, 
commute separately with   ${\rm Vir}_{\rm left}$ and  ${\rm Vir}_{\rm right}$.  As explained for example in 
\cite{Quella:2006de, Bachas:2008jd}, the topological defects  form a small subset of the much larger  class
of conformal defects and they are  characterized by a  vanishing reflection coefficient. 
 \vskip 0.6 mm  
 
 A third interesting class of defects are the {\it chiral} defects, 
 which commute with the algebra ${\rm Vir}_{\rm left}$  but not necessarily with ${\rm Vir}_{\rm right}$. 
  Chiral defects need not be scale-invariant, but the fixed points to which they flow are always
 topological. The different  classes of defects are summarized in table 1. 
 Examples of chiral defects 
  include minimal model defects perturbed by fields which are holomorphic but have fractional scaling dimension \cite{Bazhanov:1994ft, Runkel:2007wd}, 
 and  defects coupling only  to the right-moving  currents of the WZW model \cite{Bachas:2004sy}.
 In addition to ${\rm Vir}_{\rm left}$, these defects can also be shown to commute 
 with the (closed-string) Hamiltonian on the cylinder,  $L_0 + \bar L_0$.  They may thus be transported freely  in the time direction. Therefore they define conserved charges and can imprint their RG flows on boundaries.

\begin{table}
      \begin{center}
    \begin{tabular}{l | l }
   Defect type & Defining property \\ \hline
    conformal & Commutes with ${\rm Vir}_{\rm diag}$  \\
  chiral &  Commutes with ${\rm Vir}_{\rm left}$  \\
  topological  &   Commutes with ${\rm Vir}_{\rm left}\oplus{\rm Vir}_{\rm right}$\\
  holomorphic & No dependence on the left-moving sector, \\ & hence commutes with ${\cal A}_{\rm left}$
\end{tabular} 
  \end{center}
  \caption{\small The four types of defects. Holomorphic defects will be defined in section \ref{SecGlobRightSym}. ${\rm Vir}_{\rm diag}$ is the diagonal
   Virasoro algebra, while ${\cal A}$ denotes the chiral algebra, with respect to which the CFT is rational. All holomorphic defects are chiral,
   but  a chiral defect  need not be holomorphic.}
\end{table}

 
\subsection{Global versus affine group symmetries}
 
\label{ClassLoopInv}

We  specialize now to the WZW models, whose action reads \cite{Witten:1983ar}
\be\label{WZW1}
 I_{\rm WZW} =  \frac{k}{16\pi}\,  \int_\Sigma  \, {\rm Tr}^\prime  \, ( \partial^\alpha g\, \partial_\alpha g^{-1} ) 
  -   \frac{k}{24\pi}\,  \int_{\cal B}   {\rm Tr}^\prime \,  ( g^{-1} \partial_\alpha g \,  g^{-1} \partial_\beta g \, 
  g^{-1}  \partial_\gamma g )\, \epsilon^{\alpha\beta\gamma} \ ,  
\ee\noindent 
where $g$ takes values in a Lie group $G$, the level $k$ is a positive integer and ${\cal B} $
 is a $3d$-manifold whose boundary is the $2d$-spacetime $\Sigma$. To avoid heavy notation, 
 we have assumed  that $G$ is  simple and compact. More generally,  one must choose separately
  the  level of each simple factor of $G$. Following the conventions of  \cite{CFT1997}, we have
   defined  ${\rm Tr}^\prime (XY) =  {\rm tr}_R (XY)/ {x}_R$, where $x_R$ is the Dynkin
    index of the representation $R$.  The long roots of $\mathfrak{g} = {\rm Lie}(G)$ will always
     be normalized to $\sqrt{2}$. The classical field equations imply that
 \be\label{currentcons}
  \partial_\pm  {\cal J}_\mp  = 0 \  ,  \ \  \textrm{where}\ \ 
   {\cal J_-}  =   i k\,  g^{-1} \partial_- g \ \ \ \textrm{ , }\
    {\cal J_+}  =   ik\,  g \partial_+ g^{-1} \  
\ee
 and  $\zeta^\pm = \zeta^0\pm \zeta^1$.  These
are the canonically normalized currents of the WZW model which
  generate the symmetry transformations $g\, \to\, \overline \Omega (\zeta^+) \, g\,  \Omega (\zeta^-)^{-1} $,   
 i.e.  the loop extension
 of the  global $G_{\rm left}\times G_{\rm right}$ symmetry of \eqref{WZW1}.  We will denote the
 loop group by $\widehat G_{\rm left}\times \widehat G_{\rm right}$.
\vskip 1.8mm

 Consider next a generic, classically scale-invariant impurity Hamiltonian. Its associated one-form field
 \eqref{Ww1}  can be  
 parametrized   conveniently as follows: 
\be\label{gendefect}
({\cal W}_-\,  ,\,   {\cal W}_+ )   =     (\,  {\cal M}^a(g) {\cal J}^a_- \,  ,  \,    \bar {\cal M}^a(g) {\cal J}^a_+\, )  \ , \ 
\ee
where ${\cal M}^a$ and $\bar{{\cal M}}^a$ are ${\rm dim}(\mathfrak{g})$ independent matrix-valued functions on the group manifold, 
${\cal J}^a_\pm$ are the components  of the currents along the Lie-algebra direction $a$, 
and repeated indices are implicitly summed. 
The total number of independent coupling functions is therefore equal to  $2\,  {\rm dim}(\mathfrak{g})\times {\rm dim}(V)\times  {\rm dim}(V) $.
To reduce this large freedom we may impose invariance under a global   subgroup  
$H\subseteq (G_{\rm left}\times G_{\rm right})$ 
of the bulk symmetry, or under its affine extension
  $\widehat H \subseteq (\widehat G_{\rm left}\times \widehat G_{\rm right} )$.  In either case   $V$ must  carry 
a unitary $H$-representation $R$ that  describes the action of the symmetry on the defect  states,    
\be
  \Omega\in H \longrightarrow R(\Omega)\,  \in\,  {\rm End}(V)\ .  
\ee
If the symmetry is affine, the action depends on the space-time position of the defect.  Now
  the matrix elements of  
     $P e^{- \oint {\cal W}}$  will be invariant if and only if   a transformation of the bulk field transforms $ {\cal W}$  as a gauge connection: 
 \be\label{tra1}
       {\cal W}_{\alpha}   \ \to\    R(\Omega)
 \, {\cal W}_{\alpha}  \, R(\Omega)^{-1} + \,  R(\Omega)
  \partial_\alpha \  R(\Omega)^{-1}\ .  
\ee  
Of course, the inhomogeneous second term is absent  if we only require global symmetry. 
It should be stressed that ${\cal W}$ is a composite field, so its transformation is determined by that of the 
field $g$. Thus \eqref{tra1}
is  a restriction  on the couplings ${\cal M}^a$ and $\bar{{\cal M}}^a$.  As will become clear immediately, this restriction  is 
 more severe   in the affine than in the global case. 
 
\vskip 0.5mm
 
  Let us focus now on  defects    preserving  the full    left global symmetry,   
 $g(\zeta^\alpha)\to \overline \Omega g(\zeta^\alpha)$ for any constant $\overline\Omega\in G_{\rm left}$. This is a transitive symmetry,
which can be used to bring $g$ at the impurity position to any desired   value. 
Transitive global symmetries  fix all functional dependence  in ${\cal H}_{\rm imp}$ and restrict
the latter  to a finite-dimensional parameter space. 
In the case at hand, 
the covariant Hamiltonian must be given by: 
\be\label{rightS}
{\cal M}^a(g) =    - \frac{i}{k}\, \bar{R}(g)   M^a\, \bar{R}(g^{-1})
\ \ \   {\rm and}\ \ \ \  \bar {\cal M}^a(g) \,  =\,   - \frac{i}{k}\,  [{\rm Adj} (g^{-1})]^{ab}\,  \bar{R}(g)    \, \bar  M^b\,
 \bar{R}(g^{-1}) \ ,  
\ee
where $\bar{R}(g)$ is the WZW field in the representation $\bar R$ of  $G_{\rm left}$  in which the impurity states transform, 
${\rm Adj} (g)$ is the field in the adjoint representation, while $M^a$ and $\bar  M^a$ are constant  hermitean
matrices. (A factor $1/k$ has been pulled out for later convenience.) 
To verify  \eqref{tra1} one uses   $\bar{R}(\overline \Omega g)=
\bar{R}(\overline \Omega)\bar{R}(g)$ and
the simple identity
\be\label{bizzcurrent}
{\cal J}^a_+ \,  [{\rm Adj} (g^{-1})]^{ab} =  [ik\,  (\partial_+ g^{-1}) g ]^b\ .  
\ee
Both the  above expression and the right WZW currents ${\cal J}^a_-$ are  invariant under  global $G_{\rm left}$
transformations. It then  follows immediately  that   impurity Hamiltonians of the form  \eqref{rightS}
are covariant  under $G_{\rm left}$,  as advertized. 
 
\vskip 0.3mm

Can we  extend this  symmetry to $\widehat G_{\rm left}$ ?  The right  currents are invariant, but 
\eqref{bizzcurrent} transforms inhomogeneously when $\overline \Omega$ is a non-constant function of $\zeta^+$. 
 Inserting  in  the expression  for $ {\cal W}$  and
comparing   with the inhomogeneous piece in \eqref{tra1},  we deduce that the affine left symmetry
fixes $ \bar M^a = \bar T^a$, where $\bar T^a$ are the normalized generators 
of $\mathfrak{g}$ in the representation $\bar R$. 
The matrix elements of the generators coincide with those of an invariant tensor, so that 
$\ [{\rm Adj} (g^{-1})]^{ab}\, \bar{R}(g)  \,\bar T^{\, b}\, \bar{R}(g^{-1}) = \bar T^{\, a}$.  
Thus  the $\widehat G_{\rm left}$-covariant Hamiltonians take the simpler  form: 
 \be\label{genaffine}
 {\cal M}^a(g) =    - \frac{i}{k}\,   \bar{R}(g)  M^a\,  \bar{R}(g^{-1})
\ \ \   {\rm and}\ \ \ \ \bar {\cal M}^a(g)  =   - \frac{i}{k}\, \bar  T^a\  .
\ee
 The reader can verify  that any other choice for $ \bar M^a$ would fail to generate the inhomogeneous piece in (\ref{tra1})
for  non-constant  $\overline\Omega(\zeta^+)$. Notice that the covariant Hamiltonians \eqref{rightS} and 
 \eqref{genaffine}
depend on the  choice of representation $\bar R$ for the  defect states. 

\vskip 0.3mm

All $\widehat G_{\rm left}$-invariant defects are topological at the classical level. This follows from
  the field equations and the  identity 
  $\ \bar T^a {\cal J}_+^a =  ik \bar{R}(g)  \partial_+  \bar{R}(g^{-1})$,  
  which imply that  the connection ${\cal W}$ given by  \eqref{genaffine}
 is   flat for any choice of $M^a$.  
The same conclusion follows from a different argument:  the loop observables of  $\widehat G_{\rm left}$-invariant
defects,  ${\cal O}(C)$,   have  vanishing Poisson brackets with the left-moving currents which  generate this symmetry.   Since the 
 left-moving component of the  energy-momentum tensor ${\rm T}_{++}$  is quadratic in the left currents,   its  Poisson bracket 
 with ${\cal O}(C)$ is also zero.  Thus  ${\cal O}(C)$ is  chiral and,   
 being conformal,   it is automatically  topological. 
\vskip 0.7mm
 Let us pause and take stock of our main conclusion so far: The   $G_{\rm left}$-invariant   defects  of the {\rm WZW}  model with group $G$ are parametrized by 
a representation $\bar R$ of $G$  and by  $2 \, {\rm dim}(\mathfrak{g} )$ hermitean matrices $M^a, \bar M^a \in {\rm End}(V)$. For affine $\widehat G_{\rm left}$ invariance $\bar M^a$ must equal $ \bar T^a$, the generators of $G$  in the representation $\bar R$. 
\vskip 1.9mm
 
  Quantization respects the global symmetry, so it will not change the
form  \eqref{rightS} of the coupling functions. Furthermore, for the holomorphic defects studied below, the full left affine 
symmetry will be manifest since the Hamiltonian only depends on the invariant right currents. 
 Of course, because of the introduction of a UV cutoff,   conformal invariance is broken and the couplings run. 
 Nevertheless, 
 in both the global and the affine case, the  RG flow takes place in a  finite-dimensional  parameter space.  
\vskip 0.4mm

It is instructive to contrast this situation with the case of   diagonal symmetry $G_{\rm diag}$,
which maps $ g \to \Omega g \Omega^{-1}$ with  $\Omega\in G$ constant. This symmetry is not transitive
 and the general impurity Hamiltonian depends on
arbitrary functions of  the conjugacy class of $g$, i.e. of tr$(g)$. For instance
the choice  
$\ {\cal W} = -i\bar T^a [ \lambda(g)  {\cal J}_-^a \, d\zeta^- +  \bar\lambda(g)  {\cal J}_+^a\,  d\zeta^+ ]\, $  
 respects the diagonal-group symmetry for {\em any}  class functions $\lambda(g)$ and $\bar\lambda(g)$. 
 Taking these functions constant, as in ref. \cite{Bachas:2004sy}, is not however
 guaranteed by symmetry to be a stable ansatz.   The analysis of the non-chiral  defects in this reference
needs therefore to be carefully
 re-examined.\footnote{Within this restricted two-parameter space one can identify a
 (unstable)  fixed point  by imposing invariance under the affine extension of 
 $G_{\rm diag}$ which is generated by the   current combinations $\{J^a_n + \bar J^a_{-n} \vert n\in\mathbb{Z}\}$. 
 In the classical theory, the affine  $\widehat G_{\rm diag}$  symmetry requires
  $\lambda+\bar\lambda = 1/k$.  Since 
 $(\lambda, \bar\lambda) \simeq (0, 1/k)$ and $(1/k, 0)$ are stable fixed points of the RG flow, it is indeed natural to conclude that an unstable
 fixed point  lies in the middle \cite{Bachas:2004sy}.  The argument could fail at higher orders  in $1/k$, if the two-parameter restriction
 proves inconsistent. Notice that we have exchanged in this paper the roles
 of barred and non-barred couplings. }


\subsection{Holomorphic defects and their invariant subspaces}

\label{SecGlobRightSym}

The quantization of the general $G_{\rm left}$-invariant defects  \eqref{rightS}   is a very  interesting,
but technically non-trivial  problem.  What makes  it  hard, despite the huge reduction of parameter space, 
is the explicit dependence of the impurity Hamiltonian on the non-holomorphic   field $g$.  
 This difficulty persists for general  defects with  affine  $\widehat G_{\rm left}$ symmetry. Since we would like  to use the
 current-algebra method of \cite{Bachas:2004sy},  we need a Hamiltonian that only depends on the WZW currents.
 This restriction should arise from a symmetry, or else it wont be stable under RG flow. 
 Let us now examine how such a Hamiltonian can arise.
 
  \vskip 0.6mm
 
For the $g$-dependence to drop out 
of ${\cal M}^a$ we need that $M^a$ commutes with $\bar{R}(g)$ for all $a = 1 \cdots {\rm dim}(\mathfrak{g})$. For $\bar{\cal M}^a$ to be independent of $g$, we need $\bar{M}^a$ to be a $G$-invariant tensor in $\bar{R} \otimes \bar{R}^\ast \otimes \mathfrak{g}$. We can hardly be more explicit without making further assumptions on the representation $\bar{R}$. So suppose that $\bar{R}$ is a direct sum of
 $n$ isomorphic irreducible representations $\bar{r}$. Then $M^a$ has the form $M'^a \otimes \mathbb{I}_{{\rm dim}(\bar r)}$, where $M'^a$ are arbitrary hermitian matrices of size $n \times n$. On the other hand, the most general invariant tensor is of the form $\bar{M}^a = \bar{M}' \otimes \bar{t}^a$, where $\bar{M}'$ is an arbitrary $n \times n$ hermitian matrix
 and $ \bar{t}^a$ are the generators in the representation $\bar{r}$. Two special cases can occur:
\begin{itemize}
  \item If $\bar R$ is irreducible, $\bar{M}^a$ is proportional to $\bar{t}^a$ and $M^a = k m^a\mathbb{I}_n$. The defect 
is readily seen  to factorize as follows: $ {\cal O}(S^1) = e^{i m^a \int {\cal J}^a_-} \ \tilde {\cal O}(S^1)$,  
 where the first factor involves only the zero modes $\int_C d\sigma {\cal J}^a_-(\sigma)$ of the right currents and the second factor is an antiholomorphic defect depending only on ${\cal J}_+$. After quantization, the former acts like a group element on the state space of the WZW model and the latter is of the form considered in \cite{Bachas:2004sy} in the context of Kondo flows. This case therefore only leads to well known defects.
  \item If $\bar R$ is a direct sum of trivial representations, then $M^a$ are arbitrary 
  ${\rm dim}(V) \times {\rm dim}(V)$ hermitian matrices, while $\bar{M}^a$ vanishes. The impurity Hamiltonian
is in this case given by the connection one-form
\be
\label{singlet}
{\cal W}^{\rm holo} = - \frac{i}{k} \, M^a {\cal J}_-^a\  d\zeta^- \ ,     
\ee 
which now depends only on the right-moving currents. We will call this type of
 defects {\em holomorphic}.  Since  $F_{+ -}^{\rm holo} = \partial_+ {\cal W}^{\rm holo}_- =0$, holomorphic defects
 are classically topological.
  The key fact to retain here  is that the form \eqref{singlet} of  ${\cal W}$   is determined by symmetry, and 
  should  therefore remain robust  when the loop operator is renormalized. 
\end{itemize}
In general,  $\widehat G_{\rm left}$-invariant defects coupling only to the Kac-Moody currents (but not to $g$)  
  need not factorize  into a holomorphic and antiholomorphic part,
 because $M'^a$ and $\bar{M}'$ need not commute. In this paper, however, we will restrict ourselves to holomorphic defects. 
 
 \vskip 0.6mm
 
 To further reduce the parameter space of holomorphic defects we need extra symmetries. 
 These should form a global or affine subgroup of the
  remaining bulk symmetry  
 $\widehat G_{\rm right}$.  Consider first the case of a global simple subgroup  $H \subseteq G_{\rm right}$,  and let $R$ be the $H$-representation
 in which  the defect states transform.  The requirement of  $H$-invariance
 constrains  the matrix-valued vector $M$ to be a $H$-invariant tensor.
 Since the currents transform as a vector in the adjoint representation of $G$, 
 the  invariant tensors correspond to 
 equivariant embeddings  of the trivial representation of $\mathfrak{h}$ in a triple-product representation, 
\be\label{tensor}
\mathbb{C} \hookrightarrow R \otimes  R^* \otimes \mathfrak{g}  \ \equiv
\ R \otimes  R^* \otimes  ( \mathfrak{h} \oplus  \mathfrak{g}/\mathfrak{h})\   .  
\ee
Here $R^*$ is the representation conjugate to $R$,  lower-case gothic letters denote the Lie algebras,
and $\mathfrak{g}$ is considered as a (reducible for proper subgroups)  representation of $H\subseteq G$. There is one  free parameter in ${\cal H}_{\rm imp}$ for each  trivial representation  
in the decomposition of the above triple product. 
 \vskip 0.5mm

As was already discussed in the introduction, it is convenient to use an orthonormal basis   
of $\mathfrak{g}$ compatible with the decomposition $\mathfrak{g} = \mathfrak{h} \oplus \mathfrak{g}/\mathfrak{h}$. Indices $i$, $j$,... will run over
 a basis of $\mathfrak{h}$ and indices $s$, $t$,... will run over a basis of $\mathfrak{g}/\mathfrak{h}$. Splitting the adjoint vector $M^a$ accordingly, we write 
  \be\label{splitting}
  M^a {\cal J}^a_- =    \sum_{j}   \Theta^j {\cal J}^j_- + \sum_{s} \tilde\Theta^s {\cal J}^s_-\ \ .
  \ee
where $\Theta$ is an invariant tensor in $R \otimes  R^* \otimes \mathfrak{h}$ and $\tilde\Theta$ an invariant tensor in $R \otimes  R^* \otimes \mathfrak{g}/\mathfrak{h}$. 
There is a distinguished choice,  $\Theta_R$,    for the first of these tensors: it is
such that  $\Theta_R^j$ are
 the generators of $\mathfrak{h}$ in the representation $R$ with unit norm with respect to the bilinear form
 induced on $\mathfrak{h}$ from the Killing form of $\mathfrak{g}$. This distinguished tensor plays a special role when  one
 considers the extension of the global $H$ symmetry to the full affine subgroup $\widehat H \subseteq \widehat G_{\rm right}$. 
 The affine transformation $
g \to g \Omega(\zeta^-)$   implies 
 $ {\cal J}_- \to  \Omega{\cal J}_- \Omega^{-1} +  ik\,  \Omega \partial_-\Omega^{-1}\  
$, $\Omega(\zeta^-) \in H$. Using this and
 the form of
the $H$-invariant couplings  $M^a$,  gives 
  \be\label{transformH}
{\cal W}_- \  \longrightarrow  \   R(\Omega) {\cal W}_-   R(\Omega)^{-1}  + \sum_j  \Theta^j  (\Omega \partial_- \Omega^{-1})^j\ .
\ee 
The first,  homogeneous term in the transformed ${\cal W}_-$  was to be expected from the global $H$ invariance of the defect,  
while the  inhomogeneous second piece  follows from  \eqref{splitting} and  the fact that  $\Omega \partial_-\Omega^{-1}$  lies in the Lie algebra $\mathfrak{h}$.   
Comparing this transformation of ${\cal W}_-$ with the 
  required transformation  \eqref{tra1} leads to the following condition of  $\widehat H$ invariance:
 \be\label{fix}
 \Theta = \Theta_R \ \              . 
 \ee
Affine symmetry fixes, in other words, all the couplings that correspond to 
equivariant embeddings $\mathbb{C} \hookrightarrow R \otimes  R^* \otimes \mathfrak{h}$, while leaving the couplings $\tilde\Theta$ free. 
 \vskip 0.5mm

 By construction, the $\widehat H$-invariant loop operators have vanishing Poisson brackets with the currents of $H$,
\be
\{ {\cal J}_-^j \, , \ {\cal O}(C) \}_{\rm P.B.} = 0       \ . 
\ee
The condition \eqref{fix} is classical, and it is in general  modified at the quantum level. The
 safe criterion of $\widehat H$-invariance
 in the quantum theory is that the canonical commutators  which replace the above Poisson brackets vanish. 
 We will denote the invariant subspace on which this condition holds  by ${\Sigma}(\widehat H, R, k)$.

 \vskip 0.5mm
 
To summarize, invariant subspaces in the parameter space of holomorphic WZW defects can be constructed
for any choice of a subgroup $H\subseteq G$  and of  a  representation $R$.
For defects preserving the global symmetry these subspaces are
parametrized by two $H$-invariant tensors:  ${\Sigma}(H,R) = \{   \Theta ,  \tilde \Theta \}$.
Defects invariant under the affine extension belong to an invariant subspace   ${\Sigma}(\widehat H, R, k)\subset {\Sigma}(H,R)$.  
In the classical theory, this is parametrized
only by $\tilde \Theta$, since the
first set of parameters is fixed by the condition $\Theta = \Theta_R$. 
In what follows  we will often consider the case when $R$ is the restriction of an irreducible representation of $G$, with generators $T^a$.
In this case ${\Sigma}(H,R)$ contains the one-dimensional  $G$-invariant subspace $M^a = \lambda T^a$. This  intersects the $\widehat H$-invariant
subspace at a  point where  the full  
  $\widehat G$ symmetry is restored. (In 
  the classical theory the $\widehat H$-invariant subspace is   $M^j = \Theta_R^j \equiv T^j$,
and the point of  intersection  is $\lambda = 1$.)  
  
 \vskip 0.5mm  
    As we will see later,  the $\widehat G$-invariant  point is the endpoint of both the Kondo and the 
Fredenhagen-Schomerus flows. These flows take place,  respectively,  within the  $G$-invariant and
the $\widehat H$-invariant subspaces of ${\Sigma}(H,R)$.


\subsection{Examples with two parameters} 
\label{exfew} 

Let us now  illustrate the above  discussion  with some examples. Since part of our motivation
was to derive the Fredenhagen-Schomerus boundary flows as imprints of universal defect flows, 
 we will choose in our examples 
representations $R$ which are restrictions of representations of $G$. 
This  is  not in general  necessary.

 As a first example take  $G = SU(2)$ and $H= O(2)$,
with  $R$ the spin-$j$ representation of $SU(2)$.  
 The  most general $U(1)$-invariant defect Hamiltonian reads  
\be\label{su2u1}
 {\cal H}_{\rm imp}^{(j)} \ = \   \sum_{m= -j}^j   {\kappa^{(m)} }\,  \vert m \rangle \langle m\vert  
\  {\cal J}_-^3\  + \   \sum_{m= -j}^{j-1}   \left(\tilde  \kappa^{(m)}  \, \vert m+1 \rangle \langle m\vert  
\  {\cal J}_-^- \  +\   {\rm h.c.}  \right) \,  \ , 
\ee
where $\ket{m}$ is the defect state of charge $m$ in the $2j+1$ dimensional 
representation of $SU(2)$, the index $a=\pm, 3$ is the adjoint $SU(2)$ index,  and 
$({\cal J}_-^-)^* = {\cal J}_-^+$. The above Hamiltonian 
depends on  $(2j+1)$ real  parameters $\kappa^{(m)}$ and $2j$ complex  parameters $\tilde\kappa^{(m)}$.  
The minimal case $j=1/2$ has  four real parameters.
 
\vskip 0.4mm

We may further reduce the  number of free parameters in \eqref{su2u1}   
 if we impose, in addition, 
invariance under the Weyl reflection of $\mathfrak{su}(2)$. 
For $j=1/2$, in particular, the most general $O(2)$-invariant defect has just 
two real parameters, and its canonically-parametrized Hamiltonian reads\footnote{We
 have used a different symbol for these parameters, since they are not normalized as in eq. \eqref{su2u1}. Our choice of canonical normalization
 is such that the maximally-symmetric defect has $\lambda=\tilde\lambda=1$.}:
\be\label{ex2para1}
{\cal H}_{\rm imp}^{(1/2)} \ = \  - \frac{1}{k\sqrt{2}}\,   \Bigl(
\lambda \,  {\sigma^3}   {\cal J}_-^3     +\tilde  \lambda\, ( {\sigma^1} \,      {\cal J}_-^1 
+   \,   {\sigma^2 } \,     {\cal J}_-^2)  \Bigr)\ , 
\ee
where $\sigma^a$ are the Pauli matrices. In this 2-parameter space, one can distinguish three
invariant subspaces:  on $\lambda = \tilde\lambda$ the defect has global $SU(2)$ symmetry, 
on $ \lambda=1$ it has affine $\widehat {U(1)}$ symmetry, while at  the intersection
$\lambda=\tilde\lambda=1$ the full affine symmetry  $\widehat  {SU(2)}$  is  restored. 
We will revisit this example in section \ref{SecPertQuant}. 

\vskip 0.8mm
As a second example, consider $G= SU(2)\times SU(2)^\prime$ and $H$  the diagonal  $SU(2)$. 
Now ${\rm adj}({\mathfrak{g}})$ decomposes into two spin-1 representations of $H$. 
Thus, for $R = (j, 0)$ or $(0, j)$, the defect Hamiltonian has two arbitrary  parameters 
corresponding to the two trivial representations in 
the decomposition of $(j) \otimes (j) \otimes [(1) \oplus (1)]$. Explicitly, 
\be\label{ex2para2}
{\cal H}_{\rm imp}^{(j,0)} \ = \ - \frac{1}{2k} \Bigl(
\lambda \,  t^a  ({\cal J}_-^a + {\cal J}_-^{\prime a}) +
\tilde \lambda\,  t^a ({\cal J}_-^a - {\cal J}_-^{\prime a})\Bigr) \;,
\ee
with $\{t^a\}$ the generators in the spin-$j$ representation of $SU(2)$. 
The interested reader can work out the invariant subspaces in this case. 
For other representations 
of $G$  the number of free parameters rapidly  increases. 
 For instance if $R = (j , 1/2)$ with $j \geq 1$,  the tensor product
\be
\left [ \left ( j- \frac{1}{2} \right ) \oplus  \left ( j+\frac{1}{2} \right ) \right ] 
\otimes \left [ \left ( j-\frac{1}{2} \right ) \oplus \left ( j+\frac{1}{2} \right ) \right ] \otimes [(1)\oplus (1)] 
\ee
contains the trivial representation eight times. This is the number of free real parameters for defects
 with diagonal  $SU(2)$ symmetry in this model. 

\vskip 0.8mm

We conclude this subsection with a counting argument. Let  $R = r_1 \oplus \cdots r_{N_R}$ 
be the decomposition of $R$ in representations of $H$,  and $\mathfrak{g} 
= \mathfrak{h} \oplus \mathfrak{g/h} \equiv  \mathfrak{h}\oplus \mathfrak{g_1}\oplus \cdots \oplus \mathfrak{g}_{N_G}
\ $\ be a decomposition of the Lie algebra of $G$  in $H$-invariant subspaces. There exist $N_R$
independent  tensors corresponding to the generators of $\mathfrak{h}$ in the representations $r_j$, and 
at least another $N_G$ 
independent tensors that identify $\mathfrak{g}_j$ with the corresponding representation in
 $R \otimes {R}^* = {\rm adj}(\mathfrak{g}) \oplus \cdots$. Thus the number of free parameters,
 i.e. the number of equivariant embeddings \eqref{tensor}, is at least  $N_R + N_G$. Generically, this
  number  is larger. For instance, for the defects described by (\ref{su2u1}) we have $N_R = 2j+1$, $N_G = 2$, 
  while the number of free real parameters is $6j+1$.
For proper subgroups  $H\subset G$ there are at least two parameters, and 
exactly two when $R$, $\mathfrak{h}$ and $\mathfrak{g/h}$ are all irreducible under the action of $H$. 
This is precisely the form of the Hamiltonians (\ref{ex2para1}) and (\ref{ex2para2}).
Note that when counting irreducible representations  it is important to take discrete factors of $H$ into account. 

 In our discussion of RG flows, the above  2-parameter  examples will be
 simpler to analyze and to visualize, but multi-parameter cases do not present  new conceptual difficulties.


\section{Reduction to  the gauged WZW models}
\label{SecRevGaugWZW}

In this section, we will explain  how the holomorphic, $\widehat H$-invariant defects of the previous
section can be identified with local defects in the $G/H$ coset model.  
An analogous reduction to invariant sectors is well-known to work for states
in the bulk.  In order to make the paper self-contained, and to introduce some notation and conventions, 
we begin by briefly reviewing how this latter reduction works. Readers familiar with gauged WZW models may
want to skip the first two subsections and
jump directly to   \ref{LGI}.

\subsection{Review of the bulk theory}
\label{bulkrev}

The Goddard-Kent-Olive coset construction \cite{Goddard:1984vk, Goddard:1986ee} unifies in a single framework all  
 known rational conformal field theories.\footnote{However recent results \cite{Dovgard:2008zn} point to the existence of different types of rational
 conformal field theories.} This construction has been shown to have a Lagrangian description in terms
  of  the partial gauging of the $G_{\rm left}\times G_{\rm right}$ symmetry 
  of the WZW model
   \cite{Gawedzki:1988hq, Gawedzki:1988nj, Bowcock:1988xr, Karabali:1988au,1990NuPhB.329..649K, Witten:1991mm, Hwang:1993nc,Hori:1994nc, Gawedzki:2001ye}. 
Any subgroup  of the  (non-anomalous) 
diagonal symmetry, 
$H\subseteq G_{\rm diag}$, of the WZW model
can be in principle gauged by coupling the currents to a gauge connection $A$. The corresponding action reads 
 \ba\label{action}
 &&\hskip -12mm   { I}_{\rm GKO}(g, A) \,  = \,    I_{\rm WZW}(g)    +   \nonumber 
 \\ &&  \frac{k}{2\pi} \int_\Sigma {\rm Tr}^\prime \,   (A_+ g^{-1}  \partial_-g   
 +  A_-  g\partial_+g^{-1} + A_+ g^{-1} A_-  g - A_+A_-)  \ , 
\ea 
where our conventions are the same as for (\ref{WZW1}). This action is indeed invariant under the gauge transformations 
\be\label{gaugetr}
  g \to h g h^{-1} \  \  \  \ \ {\rm and} \ \ \  \ \ A_\alpha
  \to  h A_\alpha h^{-1} + h\partial_\alpha h^{-1} \ ,   
\ee
where $h(\zeta^+, \zeta^-)$ takes values in $H$, while $A_\alpha(\zeta^+, \zeta^-)$ belongs to the  subalgebra\ $\mathfrak{h} := {\rm Lie} (H)  \subseteq  \mathfrak{g}$.
 The invariance of the action follows from the Polyakov-Wiegmann identity
\be
 I_{\rm WZW}(gh) =  I_{\rm WZW}(g) +  I_{\rm WZW}(h) + 
 \frac{k}{2\pi}\,  \int_\Sigma  {\rm Tr}^\prime \,  (h \partial_+ h^{-1}\, g^{-1} \partial_- g ) \ . 
\ee
  For a detailed discussion of the gauging of the Wess-Zumino term see \cite{Witten:1991mm}, and for the effect of boundaries 
  see \cite{Gawedzki:2001ye}. Extremizing the action with respect to $A_\alpha$ gives 
\be\label{gaugeeq}
g^{-1} D_- g \Bigl\vert _{\mathfrak{h} } =   g D_+ g^{-1} \Bigl\vert_{\mathfrak{h} } = 0\ , 
\ee
where $\vert_{\mathfrak{h}}$ denotes the projection onto the Lie subalgebra $\mathfrak{h}$, and the covariant derivative is defined by $D_\alpha X:=  \partial_\alpha X + [ A_\alpha , X ]$.  Further extremizing (\ref{action}) with respect to the field $g$ gives two more equations, which for  $g^{-1} \delta g$ in $\mathfrak{h}$ or in its orthogonal complement read: 
\be\label{geq}
F(A) = 0  \hskip 3mm {\rm and}
   \hskip 3mm D_+ ( g^{-1} D_- g ) = 0 \ .  
\ee
As a check, note that the above equations reduce to those of the (ungauged) WZW model when $H$ is the identity subgroup, 
as  expected.

The action of the gauged WZW model can be rewritten in a suggestive form
 by the non-local field redefinition \cite{Gawedzki:1988hq, Gawedzki:1988nj, Bowcock:1988xr, Karabali:1988au}
\be\label{redef}
A_- :=  h_1 \partial_-  h_1^{-1}  \hskip 3mm {\rm and}
   \hskip 3mm A_+ :=   h_2 \partial_+  h_2^{-1}\  . 
\ee  
The new fields, $h_1(\zeta^+, \zeta^-), h_2(\zeta^+, \zeta^-) \in H$ are single-valued provided spacetime has no closed lightlike curves. Inserting the above expressions in the action (\ref{action}) and using the Polyakov-Wiegmann identity   gives
\be\label{action1}
 { I}_{\rm GKO}(g, A) \,  = \,   I_{\rm WZW}(h_1^{-1} g h_2)  -   I_{\rm WZW}(h_1^{-1}
 h_2 ) \ .
\ee 
The gauge transformations read: $g\to h g h^{-1}$ and $h_i\to h  h_i$, so that invariance of the action is now manifest. It also follows immediately that 
 \be\label{neweq}
 \partial_\pm {\cal J}^G_\mp = 0  \ \ \ \  \mathrm{and}\  \ \ \  \partial_\pm {\cal J}^H_\mp = 0 \ ,  
 \ee
where ${\cal J}^G_\pm$ are the WZW currents constructed from $\tilde g := h_1^{-1} g h_2
\hskip -0.6mm \in \hskip -0.6mm G$, and ${\cal J}^H_\pm$ are the currents built out of 
$\tilde h := h_1^{-1}  h_2 \hskip -0.6mm \in\hskip -0.6mm H$.\footnote{When viewed 
as a WZW action for the subgroup $H$ the second term in (\ref{action1}) has level $k^\prime = - k\,  x$, where $x$ is  the 
embedding index of $H$ in $G$. Nevertheless, we  use the normalization ${\cal J}^H_- = ik \, \tilde{h}^{-1}\partial_-\tilde{h}$, ${\cal J}^H_+ = ik \, \tilde{h}\partial_+\tilde{h}^{-1}$.} 
Because the field redefinition (\ref{redef}) involves derivatives, there exist additional non-dynamical equations, which impose constraints on
 the classical phase space. In the case at hand, they come from (\ref{gaugeeq}) and from the flatness of the gauge connection. For a cylindrical spacetime, these imply respectively 
\be\label{neweq1}
  {\cal J}^G_\pm \Bigr\vert_{\mathfrak{h} }  = {\cal J}^H_\pm \ \ \ \ 
 \mathrm{and}\  \ \ \ \ \  {\cal J}^H_\pm =  \mp  ik\, \Lambda_H  \ ,  
\ee
where $\Lambda_H$ is a constant element of the Cartan subalgebra of ${\mathfrak{h}}$. In order to derive the second equation, 
notice that in the gauge $A_+=0$, the flatness of the connection implies that $A_-$ is a function of $\zeta^-$. By a (single-valued) 
gauge transformation that depends only on $\zeta^-$, we can then bring  $A_-$ to a constant element defined up to conjugacy, 
and which can therefore be chosen in the fundamental alcove of the Cartan subalgebra. Put differently, the only physical degree of 
freedom of the gauge field on the cylinder  is a gauge-invariant Wilson line.  
\vskip 0.6mm

To summarize, the classical gauged WZW model can be described by the set of equations (\ref{gaugeeq}) and (\ref{geq}), or equivalently 
by (\ref{neweq}) and (\ref{neweq1}).  


\subsection{Quantization and state space}
\label{statespace}

Equations (\ref{neweq}) and (\ref{neweq1}) are the starting point for a canonical quantization of the gauged WZW model. 
The currents ${\cal J}^G_\mp$ form, after quantization, 
  two copies of the Kac-Moody algebra $\hat{\mathfrak{g}}_k$ at level $k$. Explicitly,  
\be\label{taylor2}
J^a_-(\sigma) = \sum_{n\in\mathbf{Z}} J^a_n\,  e^{-in\sigma} \  \ \  \mathrm{with}\ \ \ \
[ J^a_n , J^b_m ] =  i f^{abc} J^c_{n+m} + k n\,  \delta^{ab} \delta_{n+m, 0}\ ,    
\ee
and likewise for the left-moving currents $J^a_+(\sigma)$ whose modes will be denoted by $\bar J^a_{-n}$.
We work here on the cylinder, with $\zeta^1 \equiv \sigma$ the spatial coordinate. 
 The index $a$ refers to an orthonormal basis for the Lie algebra ${\mathfrak{g}}$, 
 which splits into two bases under the decomposition $\mathfrak{g} = \mathfrak{h}\oplus \mathfrak{g/h}$. 
 \vskip 0.6mm

In the canonical or GKO quantization \cite{Goddard:1984vk, Goddard:1986ee,Bowcock:1988xr} of the model, the first of the two sets of conditions (\ref{neweq1}) are imposed as operator equations. This means that the currents of $H$ are,  from the very start, identified with the naturally-embedded 
subalgebra(s) $\hat{\mathfrak{h}}_{xk} \subset \hat{\mathfrak{g}}_k$. The second set of conditions (\ref{neweq1}) can then be imposed only as 
(weak) constraints on physical states. More explicitly, if $\hat{\mathfrak{h}}_{(-)} \oplus \hat{\mathfrak{h}}_{(0)} \oplus\hat{\mathfrak{h}}_{(+)}$ is the 
Cartan-Weyl decomposition of the right-moving affine algebra, then the physical-state conditions read
\be
 J^a_n\,  \vert {\rm phys} \rangle = 0\ \ \ \ \ \ \  \mbox{for all} \;  J^a_n\in   \hat{\mathfrak{h}}_{(+)}\ \ , 
\ee  
with a similar condition for the $\bar J^a_n$. Recall that $\hat{\mathfrak{h}}_{(+)}$ contains all positive-frequency modes of the $H$-currents (i.e. all modes $J^a_n \in \hat{\mathfrak{h}}$ with $n>0$) as well as those zero-frequency generators that correspond to positive roots of the Lie algebra ${\mathfrak{h}}$.

  \vskip 0.6mm
The implementation of the above conditions amounts to decomposing the highest-weight integrable modules $L^\mathfrak{g}_{(\nu, k)}$ of $\hat{\mathfrak{g}}_k$ into $\hat{\mathfrak{h}}_{xk}$ modules:
\be
\label{DecompGMod}
L^\mathfrak{g}_{(\nu, k)} = \bigoplus_{\gamma} L^\mathfrak{h}_{(\gamma, xk)} \otimes L^{\mathfrak{g}/\mathfrak{h}}_{[\nu,\gamma]} \ .
\ee
The pairs of highest weights $[\nu,\gamma]$ label the coset fields (the level labels are here suppressed). The coset modules $L^{\mathfrak{g}/\mathfrak{h}}_{[\nu,\gamma]}$ are the graded equivalent of the branching coefficients $b_{\nu\gamma}$  in the  decomposition of representations of the corresponding Lie algebras. They carry an action of the coset vertex algebra, which contains all  normal-ordered products of generators of $\hat{\mathfrak{g}}_k$ commuting with every element in $\hat{\mathfrak{h}}_{xk}$. The modules $L^{\mathfrak{g}/\mathfrak{h}}_{[\nu,\gamma]}$ are the basic building blocks of the state space of the GKO coset models. To complete the construction one needs to mod out residual discrete gauge symmetries. This leads  to some  identifications  of coset fields -- the reader can consult \cite{CFT1997} and \cite{Hori:1994nc} for more details.

 \vskip 0.6mm

A different but equivalent approach is the BRST quantization of the theory, which was studied in \cite{1990NuPhB.329..649K,Hwang:1993nc}. In this approach one quantizes the two WZW actions of (\ref{action1}) independently, thereby obtaining two different current algebras, one  for $G$ at level $k$ and one
for $H$ at level $-xk - 2\check{h}_\mathfrak{h}$. The shift in the second level, equal to twice the dual
Coxeter number of $\mathfrak{h}$, arises from non-trivial Jacobians, which also introduce a set of decoupled ghosts. 
 We will describe the structure of the state space in the BRST formalism in the appendix \ref{BRST}. 
 
  \vskip 0.6mm
 
 The important lesson to retain from this brief review is the following: the spectrum of the $G/H$ model
 can be obtained algebraically, by first constructing the states of the associated WZW model with group $G$,
 and then projecting onto  $(\widehat H_{\rm left} \times \widehat H_{\rm right})$-invariant sectors. The auxiliary WZW fields, 
 on the other hand,  are related to the local fields, 
 $g$ and $A_\alpha$,    by the non-local   redefinition
\eqref{redef} and  \eqref{action1}.  To prove that the $\widehat H$-invariant defects of subsection 
 \ref{SecGlobRightSym} can be identified with GKO defects, we need to work backwards, i.e. to  show that they  
arise from  local gauge-invariant  couplings  to $g$ and $A_\alpha$. 
   
  
\subsection{Gauge invariant defects}
\label{LGI}    

For the WZW defects studied in  section  \eqref{SecDefLoop} the imposition of (global or affine)
symmetries was optional. In a gauge theory, on the other hand, only gauge-invariant probes 
are allowed. Thus, if  ${\cal W}(g, A)$ is the composite connection form  integrated along the defect loop, 
then under the gauge transformations  \eqref{gaugetr}  we must  have
 \be\label{tra2}
       {\cal W}_{\alpha}   \ \to\    R(h)
 \, {\cal W}_{\alpha}  \, R(h)^{-1} + \,  R(h)
  \partial_\alpha \  R(h^{-1})\ ,   
\ee  
where the defect transforms in a representation $R$ of the gauge group $H$.  
Condition \eqref{tra2}  is similar to the condition \eqref{tra1} of section \ref{SecDefLoop}, with one  
important  difference:  the transformations $\Omega$  in  section \ref{SecDefLoop} were elements
of the loop group, whereas  here $h(\zeta^0, \zeta^1)$ can have arbitrary  dependence on the spacetime coordinates. 

\vskip 0.5mm

The simplest choice obeying \eqref{tra2} is 
${\cal W}_{\alpha}  = \sum_{j}  \Theta_{R}^j A^j_\alpha\, $,
with
 $\Theta_{R}^j$ the generators of $H$ in the representation $R$
(see section \ref{SecGlobRightSym}). This choice corresponds to the standard
 Wilson loop of  an external probe coupling minimally
to the gauge field $A_\alpha$. More general couplings are however possible.  Any extra term which is of dimension one (for classical scale invariance) and transforms
homogeneously is suitable. It is easy to construct such allowed  couplings  using the covariant
``currents",   $g^{-1}D_\alpha g$, and class functions.  We will not try
here to be exhaustive, but rather  focus immediately on the gauge-invariant defects that will make
contact with the holomorphic  defects of WZW models. These correspond to the choice
\be
\label{flatc}
\mathcal{W}  \ =\   \sum_{j}   \Theta_R^j A^j_\alpha\,  d\zeta^\alpha + \sum_{a}\tilde \Theta^a  \,  (g^{-1} D_- g)^a \, d\zeta^-\ , 
\ee
where $\tilde\Theta^a$ are the components of an $\mathfrak{h}$-invariant tensor on $\mathfrak{g} \otimes R \otimes R^\ast$. 
 A simple calculation gives for the field strength of the above connection form 
\be
\epsilon^{\alpha\beta} 
\left( \partial_\alpha \mathcal{W}_\beta  - \partial_\beta \mathcal{W}_\alpha + [ \mathcal{W}_\alpha , \mathcal{W}_\beta ] \right)\, 
= \, \sum_{j} \Theta_{R}^j F(A)^j  +    \sum_{a} \tilde \Theta^a   (D_+ ( g^{-1} D_- g ))^a \ . 
\ee
By virtue of the field equations (\ref{geq}), the right-hand side vanishes so the connection \eqref{flatc} is flat. The
corresponding defects are therefore classically topological. 

\vskip 0.5mm

 Since the loop operators of the above defects are gauge invariant, we can evaluate them in any given gauge.
 A convenient choice is $h_2 = 1 \Longrightarrow A_+=0$, where  $h_2$ was defined by eq.  \eqref{redef}. 
A straightforward calculation, using the definitions of the auxiliary WZW currents  given  in subsection \ref{bulkrev}, then leads to  $\mathcal{W}_+ = 0$ and 
 \be\label{flatcc}
i k\,  \mathcal{W}_- \  =\    \sum_{j} \left[  \Theta_{R}^j ({\cal J}^H_-)^j + \tilde \Theta^j ({\cal J}^G_- - {\cal J}^H_-)^j  \right] \,  + \,
  \sum_{s} \tilde \Theta^s ({\cal J}^G_-)^s  \ . 
\ee 
Notice that in this gauge  the defect  loop can be expressed entirely in terms of the right-moving auxiliary  WZW currents.
The second term in the 
square brackets vanishes, at the classical level, because of  the current
 identification \eqref{neweq1}. In the canonical (GKO)  quantization,  this identification holds as an  operator identity so we may as well consider the simpler connection
  \be\label{flatccc}
i k\,  \mathcal{W}_- \  =\    \sum_{j}  \Theta_{R}^j ({\cal J}^G_-)^j  \,  + \,
  \sum_{s} \tilde \Theta^s ({\cal J}^G_-)^s  \ . 
\ee 
In the BRST quantization of the coset model, on the other hand, one has to work with the form  \eqref{flatcc} of the gauge connection. 
 \vskip 0.5mm
 
 Equation  \eqref{flatccc} is  the main result of this section.
It shows that,  when expressed in terms of the auxiliary WZW currents, our  class of GKO  defects  is
the same as the $\widehat H$-invariant holomorphic defects analyzed in section 2. 
This follows from the comparison of the above expression with eqs. 
 \eqref{singlet},  \eqref{splitting}   and \eqref{fix} of section 2. 
 Notice, in particular, that the defect coupling 
to the   currents of $\mathfrak{h}$ has been  frozen precisely as in eq.  \eqref{fix}. 
The couplings $\tilde\Theta^j$  are unphysical and have dropped out from the final expression for $ \mathcal{W}$. 
 This can  be  also seen more directly from the covariant eqs. \eqref{flatc} and \eqref{gaugeeq}.
 \vskip 0.5mm
 
  The above identification  of   WZW and  GKO defects holds at the classical level. In the quantum theory 
both terms in  \eqref{flatccc} are renormalized and the invariant subspace of parameter space is
implicitly   defined
by the conditions 
\be\label{ward}
[  {J}_-^j\,  , \,  {O}(C) ] = 0 \ ,  
\ee
for all generators $ {J}_-^j$ of $\widehat{H}$. 
These conditions must be imposed order  by order in the $1/k$  expansion.  Since
the affine $\widehat H$  symmetry   cannot have an anomaly 
on the one-dimensional  worldline of the defect, we expect no obstruction to 
imposing this gauge symmetry at the quantum level. Recall that in the WZW model  the parameter subspace 
  ${\Sigma}(\widehat H, R, k)$ 
was   fixed by the requirement of  $\widehat G_{\rm left} \times \widehat H_{\rm right}$ symmetry
for defects transforming in the $(0, R)$ representation of the symmetry groups. 
Based on the above identification, one would
then conclude that such an invariant subspace of  parameter space also exists for
 the GKO defects. 
  This would have been hard to show directly, because the GKO model has no Kac-Moody symmetries.
   The existence of holomorphic defects
  in the  latter model  means that the left-Virasoro and the gauge symmetries  can be  compatibly imposed
  at the quantum level.\footnote{The affine $\hat H$ symmetry can be identified with a residual gauge symmetry, if one
  chooses the  most general gauge condition 
   consistent with $A_+ = 0$, i.e.   $h_2 = f(\zeta^-)$ for arbitrary $H$-valued function $f$.  Eq. \eqref{ward}
   can then be interpreted as the  Ward identity which guarantees that the choice of $f$ should not matter.}
 \vskip 0.5mm

The Wilson loop of the connection \eqref{flatccc} has a simple interpretation at the special value $\tilde\Theta^s = 0$
and, when $R$ is the restriction of a representation of $G$, also at the special value $\tilde\Theta^s = T^s$ (with $T^a$ the
generators of $G$ in $R$).  At these special values, it measures the classical monodromies of the gauge-invariant fields 
$\tilde h$ and $\tilde g$ defined in subsection  \eqref{bulkrev}. These fields obey the WZW equations, 
 which imply the following factorization into left-moving and right-moving parts:  
\be
\tilde h(\zeta^+ , \zeta^-) =   \tilde h_+^{-1}(\zeta^+)\,  \tilde h_-(\zeta^-) \ \ \  \mathrm{and} \ \ \ 
\tilde g(\zeta^+ , \zeta^-) =  \tilde g_+^{-1} (\zeta^+)\,  \tilde g_-(\zeta^-) \ . 
\ee
Solutions on the cylinder are therefore classified by their classical monodromies
\be
\tilde h_\pm (\zeta^\pm \pm 2\pi) =  u^H \,  \tilde h_\pm (\zeta^\pm) \ \ \  \mathrm{and} \ \ \ 
\tilde g_\pm (\zeta^\pm \pm 2\pi) =  u^G\,  \tilde g_\pm (\zeta^\pm)\ , 
\ee
where $u^H\in H$ and $u^G\in G$ are constant group elements.  The above loop observables
are   traces of these  constant  matrices in the representation $R$, 
\be
{\cal O}(S^1) = \begin{cases}  \mathrm{tr}_R( {u}^H) & \text{if $\tilde\Theta^s = 0$,}
\\
\mathrm{tr}_R ({u}^G)  &\text{if $\tilde\Theta^s = T^s$.}
\end{cases}
\ee
The values of these traces determine the conjugacy class of the monodromies.
This is the only non-ambiguous data, since $\tilde h_\pm$  and $\tilde g_\pm$ can be redefined by left multiplication with constant group elements. Using this freedom one can bring the monodromy matrices to canonical form: 
\be
 {u}^H = \exp (2\pi \Lambda_H)  \ \ \  \mathrm{and} \ \ \  {u}^G = \exp (2\pi  \Lambda_G) \ ,  
\ee
where $\Lambda_H$ and $\Lambda_G$ belong to the corresponding Cartan tori. The classical monodromies take continuous values in these tori, while in the quantum theory they are discretized as we will later see. In the special case of the abelian
 WZW model, i.e. of a free compact scalar field, ${\Lambda}_G$ is just 
 the momentum zero mode.  
\vskip 1mm

To summarize, we have identified a family of flat gauge connections in the gauged WZW models.
 In a specific gauge, they can be expressed in terms of the right-moving auxiliary current $\mathcal{J}^G_-$, 
 and can therefore be studied within the ungauged WZW model. The corresponding Wilson loops
  define gauge-invariant, classically-conserved observables,  which interpolate between traces of classical monodromies at 
   the two special points $\tilde\Theta^s= 0$ and $\tilde\Theta^s = T^s$.  
 Analogous observables can be, of course, constructed with  the left-moving currents. 
 We turn now to the quantization of these classical observables. As we will show, the classical-monodromy points 
 will correspond to fixed points of the renormalization group flow.


\section{Perturbative quantization and RG flows}
\label{SecPertQuant}

In  this section we will  quantize perturbatively in ${1}/{k}$ the holomorphic
WZW defects \eqref{singlet} and  study their renormalization-group flows.  This is an extension
of  the analysis of refs.   \cite{Bachas:2004sy, Alekseev:2007in}. The main result is  summarized by the RG flow
equation  \eqref{BetaFun}.   
We will   verify to the leading order, and argue more
generally, that the   flow preserves the gauge-invariant subspaces of parameter space
described in the previous subsection. On these subspaces, the defects and RG flows of the WZW model
can be identified with defects and flows of the corresponding coset model.

\subsection{Regularization of loop operators}
\label{SecRegLoop}

As has been shown in  section \ref{SecDefLoop}, the most general 
 holomorphic, $\hat{G}_{\rm left}$-invariant defect of the WZW model  is parametrized by  dim$(\mathfrak{g})$ hermitean
 matrices $M^a$, so that  ${\cal H}_{\rm imp}^{\rm holo} = -\frac{1}{k} \, M^a {\cal J}_-^a\ $. We will
 occasionally drop the adjoint index, and write $M$ for the vector of matrices whose components are $\{M^a\}$.
  We work on the cylinder with 
  coordinates $(\zeta^0, \zeta^1) = (\tau, \sigma)$ where $\sigma \simeq \sigma +2\pi$ is the periodic coordinate.
  Throughout this section $C= S^1$ will be  the $\sigma$ circle, and the dependence of the loop
  observables on $C$ will be dropped.   
  (Note that with our conventions the "Hamiltonian" for a spacelike curve is\ \  $i{\cal H}_{\rm imp}^{\rm holo} = {\cal W}^{\rm holo} _-$). 
  Taylor expanding the exponential in \eqref{FormLoopOp} gives the following expression for the loop operator:
 \be
\label{taylora}
  {\cal O}(M) = \sum_{N=0}^\infty\, (ik)^{-N}\, \mathrm{tr} (M^{a_1} \cdots M^{a_N})  
\oint_{\sigma_1}\hskip -1.4mm  \cdots  \hskip -1mm \oint_{\sigma_N} \hskip -1mm \theta({\sigma_1\geq \cdots \geq \sigma_N}) 
 \,  {\cal J}_-^{a_1} (\sigma_1)
\cdots {\cal J}_-^{a_N} (\sigma_N)\ .
\ee
Here the path-ordering function $\theta$ equals one
 when $2\pi \geq \sigma_1\geq  \cdots \geq \sigma_N\geq 0$ and zero otherwise, and ${\cal J}^a_-(\sigma)$ are the right-moving classical currents. 
 Notice that the above Taylor expansion  is written as a series in inverse powers of $k$. 
 To justify a perturbative treatment, we will assume that the level $k \gg 1$,\footnote{This is  the semiclassical limit of the WZW model, 
 in which the  target-space  curvature  is  small.}
 and  that the  coupling matrices $M^a$ have entries of order 1. 
\vskip 0.5mm

In order to quantize the operator \eqref{taylora} we must replace the classical currents
by their quantum counterparts. We will use  calligraphic and regular symbols to distinguish these two. 
The quantum currents  generate  a Kac-Moody algebra: 
     \be\label{taylorb}
J^a_-(\sigma) = \sum_{n\in\mathbf{Z}} J^a_n\,  e^{-in\sigma} \  \ \  \mathrm{with}\ \ \ \
[ J^a_n , J^b_m ] =  i f^{abc} J^c_{n+m} + k n\,  \delta^{ab} \delta_{n+m, 0}\ .    
\ee
Inserting this in \eqref{taylora} gives a divergent expression, because products of quantum currents at coincident
points are singular.  We can remedy the situation by introducing
 a frequency cutoff\,: 
\be
\label{taylor3}
J^{a}_- (\sigma)
\to J^a_{{\rm reg}} (\sigma) :=  \sum_{n\in\mathbf{Z}} J^a_n\,  e^{-in\sigma -\vert n\vert \epsilon/2 } \;, \;\; \epsilon > 0\ .  
\ee
 This makes operator products finite, but 
 does not fully specify the regularization prescription, because of a subtle and important  issue which we now discuss.   
  
 \vskip 0.5mm

The issue has to do with operator ordering, and with the symmetry of the loop operator under (rigid) translations
of $\sigma$.  
In the classical theory the order of multiplication of the currents in  eq. (\ref{taylora}) does not matter
(only the order of multiplication of the matrices $M^{a_i}$ does). 
The same would be true in the quantum theory if the singularities at $\sigma_j = \sigma_{j+1}$ were regularized by point splitting, because the currents $J^{a}_- (\sigma)$ at space-like separations commute. However the regularized currents \eqref{taylor3} do not commute, even when $\sigma_i \not= \sigma_j$. Thus, as part
of the regularization prescription for the loop operator, 
 an ordering of the r.h.s. in eq.  (\ref{taylora}) must be specified. A good choice consists in summing over the $N$ cyclic permutations of the currents,
 and over the $N$ order-reversed permutations for which one must furthermore complex-conjugate the matrices $M^a$. The full
 regularization prescription reads: 
 \begin{align}
\label{replace}
\mathrm{tr} (M^{a_1} \cdots M^{a_N}) &\mathcal{J}^{a_1}_- (\sigma_1) \cdots  \mathcal{J}^{a_N}_- (\sigma_N) \to  \notag \\ 
&\frac{1}{2N} \mathrm{tr} (M_{\rm o}^{a_1}  \cdots M_{\rm o}^{a_N}) \sum_{\pi \in \mathbb{Z}/N\mathbb{Z}}  
J^{a_1}_{\rm reg} (\sigma_{\pi(1)}) \cdots  J^{a_N}_{\rm reg} (\sigma_{\pi(N)}) \; + \\
&\frac{1}{2N} \mathrm{tr} ((M_{\rm o}^{a_1})^\ast \cdots (M_{\rm o}^{a_N})^\ast ) \sum_{\pi \in \mathbb{Z}/N\mathbb{Z}} 
J^{a_N}_{\rm reg} (\sigma_{\pi(N)}) \cdots  J^{a_1}_{\rm reg} (\sigma_{\pi(1)}) \notag \;.
\end{align}
We have here introduced the symbol $M_{\rm o}$ for the {\it bare} coupling matrices,  while $M$
will stand for the renormalized ones.  
As explained in \cite{Bachas:2004sy}, the above regularization respects the classical symmetries  under translations of $\sigma$, 
as well as under hermitian conjugation followed by orientation reversal.    The $\sigma$-translations  are 
generated by the combination $L_0-\bar{L}_0$ of the Virasoro zero modes. Being holomorphic, the regularized loop operator 
furthermore commutes with all  left Virasoro generators. It is thus also invariant under translations along $\tau$, 
generated by $L_0 + \bar{L}_0$. This crucial property ensures the universality of induced boundary flows (cf. \cite{Bachas:2004sy}). 

\vskip 0.5mm 

Making the replacement (\ref{replace}) in the expansion (\ref{taylora}), inserting the  mode expansion (\ref{taylor3}) and performing the $\sigma$ integrations, 
gives  a well-defined regularized expression,  $O_{\rm reg}(M_{\rm o}, \epsilon)$,  for the loop operator 
as a series in the Laurent modes $J^a_n$. 
To compute the action of this operator on the highest-weight modules
 forming the state space of the theory, we have to normal order the currents. 
 This  produces terms that diverge when $\epsilon\to 0$. 
 Our main (renormalizability) hypothesis is that  these divergences can be absorbed  in a redefinition of the couplings,
 and in an overall multiplicative factor.  More precisely, we assume that there exist  effective couplings
  $M(M_{\rm o},\epsilon)$,  and a constant  (c-number) multiplicative factor  $Z(M_{\rm o},\epsilon)$ such that the following limit  
  is well-defined:  
\be\label{renormlimit}
O_{\rm ren}(M) := \lim_{\epsilon\to 0}\      \   Z(M_{\rm o},\epsilon)  \, O_{\rm reg}(M_{\rm o}(M, \epsilon) , \epsilon ) \ . 
\ee
Here  -log$Z$ is a  self-energy counterterm, and $M_{\rm o}(M, \epsilon)$ is the inverse of the matrix-valued
function $M(M_{\rm o},\epsilon)$. This inversion is possible at least near $M_{\rm o} = 0$,  where $M = M_{\rm o} + o(M_{\rm o}^2)$. 
The existence of the limit  \eqref{renormlimit} is based on the assumption of power-counting renormalizability, which states
that all divergences can be removed by subtracting from ${\cal H}_{\rm imp}$ local counterterms of dimension $\leq 1$. 
 The only such counterterms consistent with the $\hat{G}_{\rm left}$ symmetry is a constant\footnote{In string theory 
this is a tachyon background, which must be constant because ${G}_{\rm left}$ acts transitively. Note that  log$Z$ is a real
counterterm, while the renormalized evolution operator is unitary.  This is because the defect coupling to the currents, 
like   higher-dimensional  Chern-Simons terms, acquires an $i$ upon Wick rotation.  }  
plus couplings of the same form 
as  \eqref{singlet}. 
 
\vskip 0.5mm   
 
 We will not give here a proof of renormalizability at all orders in the perturbative expansion. This is an
 interesting mathematical problem, but we expect no surprises. 
 We will limit ourselves to a computation of the leading $1/k$ contribution to $M(M_{\rm o},\epsilon)$,  
 and of  the associated $\beta$-function that describes the evolution of the couplings from the ultraviolet to the infrared.


\subsection{Renormalization and the $\beta$-function}

\label{SecRenBeta}

Let us consider  the $N$th term in the Taylor expansion (\ref{taylora}) and denote the corresponding
 operator, regularized as was explained in the previous subsection,  by  $(ik)^{-N} O_{\rm reg}^{(N)}$. A tedious but straightforward calculation \cite{Bachas:2004sy}
 gives the following expressions for the first four operators of the Taylor series in terms of the Kac-Moody currents:
\begin{equation}\label{R2}
  O_{\rm reg}^{(1)}   = 2\pi\, {\rm tr} (M_{\rm o}^a)\;  J_0^a\ , \ \ \ \ O_{\rm reg}^{(2)}   = 2\pi^2 \, {\rm tr} (M_{\rm o}^a M_{\rm o}^b)\;  J_0^a J_0^b\ ,
\end{equation}
\vskip -1mm
\be 
\label{R3}
O_{\rm reg}^{(3)} = \frac{2\pi^2}{3} \; {\rm tr} (M_{\rm o}^a M_{\rm o}^b M_{\rm o}^c) \Bigl[ \, \frac{\pi}{3}\; J_0^a  J_0^b J_0^c\ + \sum_{n\not= 0 }
 \frac{i}{n} J^a_{-n} J^b_{n} J^c_0\;  e^{-\vert n\vert \epsilon}  + {\rm cyclic} + {\rm reversed}\; \Bigr]\ ,
\ee
and
\begin{eqnarray}\label{R4} 
\hskip -0.1cm O_{\rm reg}^{(4)} \hskip -1mm &=& \hskip 1mm  \frac{\pi^2}{2} \; {\rm tr}(M_{\rm o}^a M_{\rm o}^b M_{\rm o}^c M_{\rm o}^d)\; \Biggl[\, \frac{\pi^2}{6} 
 J_0^a  J_0^b  J_0^c J^d_0 \,+ \,\sum_{n\not= 0} \frac{i\pi}{n} J^a_{-n}  J^b_{n} J^c_0  J^d_0 \, e^{-\vert n\vert \epsilon} \nonumber \\
&&\hskip -1.5cm  \quad + \sum_{n\not= 0 } \frac{1}{n^2} \left(J^a_{-n}  J^b_{n} J^c_0  J^d_0 -  J^a_{-n} J^b_{0} J^c_n  J^d_0\right) e^{-\vert n\vert \epsilon} + \hskip -2mm \sum_{\stackrel{\scriptstyle m,l,n\not=0}{m+n+l=0}} \hskip -1mm  \frac{1}{ml}\,  J^a_{m}  J^b_{n} J^c_l  J^d_0\,   e^{-(\vert n\vert +\vert m\vert +\vert l\vert ) \epsilon/2 }    \nonumber  \\
&& \quad - \frac{1}{2} \sum_{m,n\not= 0}\frac{1}{mn} J^a_{-n} J^b_{n} J^c_{-m} J^d_m \  e^{-\vert n\vert \epsilon -\vert m\vert \epsilon} \; +\; {\rm cyclic} + {\rm reversed} \;   \Biggr]\ .  
\end{eqnarray}
Repeated upper indices in these expressions are implicitly summed, and  ``cyclic'' or ``reversed'' denote the terms generated by cyclic or reversed permutations, 
see \eqref{replace}.  For reasons that   will become clear in a minute, the
 operators $O_{\rm reg}^{(N)}$ for $N\geq 5$ will not be needed  in our leading-order calculation of the $\beta$-function. 

\vskip 0.5mm

  In order to organize the calculation, note first that without loss of generality we can take all matrices $M^a$ (bare and renormalized)  to be traceless. 
Indeed, any  $M^a$ can be written as  
$M^a = m^a k  \mathbb{I} + \tilde M^a$,  with $m^a$  a number and  $\tilde M^a$  traceless.   The term
proportional to the identity  commutes 
(in the classical theory) with everybody else. It  can be pulled outside the path-ordering prescription, so that  
 the loop observable factorizes:  ${\cal O}(M) = {\rm exp}(  i m^a \oint {\cal J}^a_-)\,  {\cal O}(\tilde M)$. 
The first factor  depends only on the current zero modes, and  requires no renormalization. 
We may thus drop it (as well as the tilde) and take $M^a$ traceless. 
In this case  $O_{\rm reg}^{(1)} = 0$ and the first non-trivial term 
 in the Taylor expansion is $O_{\rm reg}^{(2)}$.

\vskip 0.5mm

 Let us focus on the divergent contributions to  this term, which arise from the normal ordering of   $O_{\rm reg}^{(N)}$
 for $N\geq 3$.  We want to regroup these corrections into a single  expression: 
\be
\label{QuadTermRenWL}
-\frac{2\pi^2}{k^2}\,  {\rm tr} \left ( M^a M^b \right ) J^a_0 J^b_0 \ , \ \ \ {\rm with}\ \ \  M^a = M_{\rm o}^a + {1\over k} \Delta_1^a (M_{\rm o}, \epsilon) + 
{1\over k^2} \Delta_2^a (M_{\rm o}, \epsilon) + \cdots 
\ee
The leading correction to the effective coupling,  $\Delta_1$,  can only arise from the $N=3,4$ terms in the Taylor
expansion of the loop operator.  This is because the $N$th term  starts out with $N-2$ more currents and more powers 
of $1/k$ than the $N=2$  term.   Reordering these currents can at most compensate $[N/2 - 1]$ inverse powers of $k$. Here
$[x]$  stands for the integer part of $x$,  and the maximal compensation arises when all current commutators are 
replaced by the central term of the Kac-Moody commutator. It follows that   $O_{\rm reg}^{(N)}$
can   contribute  to the $n$th correction of the effective coupling in the quadratic  term
only  if  $ N - 2 \geq  n \geq N - 2 - [N/2-1]$.  For $n=1$ this gives $N=3$ or $4$,  as claimed. 
Moreover,  $O_{\rm reg}^{(4)}$ can  only  contribute if  two  currents  are replaced by the central term of the Kac-Moody commutator.
These remarks  simplify drastically the calculation of $\Delta_1^a (M_{\rm o}, \epsilon)$.
\vskip 0.5mm

To perform this calculation, we must normal-order the operators $O_{\rm reg}^{(N)}$, i.e. pass all the 
positive-frequency modes to the right of all the negative modes. The prescription is not unique because (unlike what happens
for the free field) same-sign current modes do not commute.  The rearrangement  of same-sign modes does not, however, introduce new divergences. 
Following  \cite{Bachas:2004sy}, we define the normal-ordered cubic operator as follows: 
\begin{eqnarray}
:O^{(3)}(M): &=& \frac{4\pi^3}{3} \; {\rm tr} (M^{\{ a}
M^b M^{c\}} ) \;   J_0^a  J_0^b J_0^c\  \nonumber\\
&&\hskip 1cm   -  \,  {4\pi^2} \; {\rm tr} ([M^a , M^b] M^c) \, 
\sum_{n >0 } \frac{i}{n}  J^a_{-n}  J^b_{0}
J^c_n\  , 
\end{eqnarray} 
where curly brackets denote symmetrization of the indices.  A straightforward calculation,
with the help of  the Kac-Moody commutators,  then gives: 
\begin{eqnarray}\label{03no}
O_{\rm reg}^{(3)}(M_{\rm o}) &=&   : O(M_{\rm o}):   
    -\,  4\pi^2\,  {\rm tr} ( [M_{\rm o}^a,  M_{\rm o}^b] M_{\rm o}^c) f^{abd} \sum_{n >0 } \frac{1}{n} ( J^c_{-n}  
J^d_n + J^d_{-n}  
J^c_n
) \nonumber \\
&& \hskip 1cm +\,  2 \pi^2\,  {\rm tr} ([M_{\rm o}^a,  M_{\rm o}^b] M_{\rm o}^c) f^{abd} \,  J^c_0 J^d_0 \left( \sum_{n>0} \frac{e^{-\epsilon n}}{n} \right)
 \nonumber \\ && \hskip 1cm   -\, 
\frac{2\pi^2}{3} \; k\,  {\rm tr} ([M_{\rm o}^a,  M_{\rm o}^b]  M_{\rm o}^c) f^{abc}\,  \left( \sum_{n>0} {e^{-\epsilon n} } \right) \ ,  
 \end{eqnarray} 
up to terms that vanish in the limit $\epsilon\to 0$. The first line in the above expression is a well-defined operator,
equal to the normal-ordered product plus  a finite correction. The second line is a logarithmically-divergent
addition to  $O^{(2)}$,  which we will   absorb  in a redefinition of the couplings $M^a$.  Finally, the third
line gives a (linearly-divergent) self-energy correction that can be absorbed in the multiplicative constant $Z(M_{\rm o}, \epsilon)$. 

\vskip 0.5mm

  For the $N=4$ term of  the Taylor expansion we will be less systematic, and focus directly on the divergent, order $k$ 
   contributions to the quadratic operator $J_0^c J_0^d$.  More explicitly, we write
  \be\label{xi}
  O_{\rm reg}^{(4)}(M_{\rm o}) \ =\   : O^{(4)}(M_{\rm o}):   \ + \  k\, \Xi^{cd}(M_{\rm o}, \epsilon)  \, J_0^c J_0^d +  \cdots \ , 
  \ee
where the dots include contributions to the quadratic operator of order $k^0=1$,  
renormalizations of the matrices $M^a$  entering  in the cubic operator,   self-energy corrections,  and finite terms. 
Notice that $ O_{\rm reg}^{(4)}(M_{\rm o})$ comes multiplied by  a factor   $k^{-4}$, so that the $O(1)$ contributions
to the quadratic operator  renormalize the couplings $M$ at order $k^{-2}$. Since we are here only interested
in the  renormalization of order $k^{-1}$,  such contributions can be safely dropped.  Note also that,  by 
the assumption of renormalizability, we only need to compute the corrections to  $J_0^a J_0^b$.
All the other operators should be made finite by the {\it same} redefinition of the couplings $M$. 

\vskip 0.5mm

   To calculate $\Xi^{ab}$ we will  consider  each term of  $O_{\rm reg}^{(4)}(M_{\rm o})$,   as given 
  by eq.  \eqref{R4}, in turn.   Recall that we need to replace one commutator by the central extension of the
  Kac-Moody algebra, in order to get the desired factor of $k$. We use the notation $X \hookrightarrow \xi^{cd}$  
  to mean that,  after normal ordering, the operator $X$ makes a contribution $\xi^{cd}$ to $k \Xi^{cd}$. 
  We then have:
  
 \begin{itemize}
	\item $J_0^a  J_0^b  J_0^c J^d_0$ : This term needs no normal ordering and does not generate a contribution to $\Xi^{cd}$.
\item ${1\over n}\, J^a_{-n}  J^b_{n} J^c_0  J^d_0$ : This term could contribute  through the central term in  the commutator  $[J^a_{-n} , J^b_{n}]$.  
A careful calculation however gives: 
	$$
	\frac{i\pi^3}{2}\;{\rm tr}(M_{\rm o}^a M_{\rm o}^b M_{\rm o}^c M_{\rm o}^d) \sum_{n \neq 0} ( \frac{1}{n} J^a_{-n} J^b_n J^c_0 J^d_0 e^{-|n|\epsilon} + \; {\rm cycl.} \; + \; {\rm rev.}  )   
	$$
	$$
	\hookrightarrow\    2i\pi^3 k \;{\rm tr}(M_{\rm o}^a M_{\rm o}^a M_{\rm o}^c M_{\rm o}^d)   \left ( \sum_{n>0} e^{-|n|\epsilon} - \sum_{n<0} e^{-|n|\epsilon} \right ) = 0 \;.
	$$
We used here the cyclic invariance of the trace, and  the fact that the  matrices $M_{\rm o}^a$ are hermitean.
 The cancellation between the  $n >0$ and $n<0$ sums 
 is due to  the factor of $n$ in the central extension: depending on the sign of $n$, either $J_{n}J_{-n}$ or $J_{-n}J_n$  must be reordered.
 Notice that this  contribution would have been linearly divergent, so by power-counting it could not possibly renormalize the
 couplings $M$. 
   \item ${1\over n^2}\, J^a_{-n}  J^b_{n} J^c_0  J^d_0$ : Because of the extra factor of $n$, the (now logarithmically-divergent) contributions  do
   not cancel.  A straightforward calculation gives: 
 $$
	\frac{\pi^2}{2} \;{\rm tr}(M_{\rm o}^a M_{\rm o}^b M_{\rm o}^c M_{\rm o}^d) \sum_{n \neq 0} ( \frac{1}{n^2} J^a_{-n} J^b_n J^c_0 J^d_0\,  e^{-|n|\epsilon} + \; {\rm cycl.} \; + \; {\rm rev.} ) 
	$$
 \be\label{xi1}
	\hookrightarrow\ 4\pi^2 k \;{\rm tr}(M_{\rm o}^a M_{\rm o}^a M_{\rm o}^c M_{\rm o}^d)  \left( \sum_{n > 0} \frac{e^{-|n|\epsilon}}{n}  \right) \;.
\ee
\item $ {1\over n^2}\,  J^a_{-n} J^b_{0} J^c_n  J^d_0$ : This term makes a similar contribution. After renaming dummy  indices: 
 	$$
	-\frac{\pi^2}{2} \;{\rm tr}(M_{\rm o}^a M_{\rm o}^c M_{\rm o}^b M_{\rm o}^d) \sum_{n \neq 0} ( \frac{1}{n^2} J^a_{-n} J^c_0 J^b_n J^d_0\,  e^{-|n|\epsilon} + \; {\rm cycl.} \; + \; {\rm rev.})
	$$
\be\label{xi2}
	\hookrightarrow\  -4\pi^2 k \;{\rm tr}(M_{\rm o}^a M_{\rm o}^c M_{\rm o}^a M_{\rm o}^d)  \left(  \sum_{n > 0} \frac{e^{-|n|\epsilon}}{n}  \right) \;.
\ee
 \item $ {1\over ml}\, J^a_{m}  J^b_{n} J^c_l  J^d_0$  :  This does not contribute to $\Xi^{ab}$, because it contains no pair of currents
 with equal but opposite frequencies, whose commutator could give the central term. 
 \item ${1\over nm}\,  J^a_{-n} J^b_{n} J^c_{-m} J^d_m$ :  Likewise it does not contribute  because $n, m\not= 0$, so there is no
 way to produce the two necessary  current zero modes.
 \end{itemize}

To summarize, $k\Xi^{ab}$ is the sum of  the two expressions \eqref{xi1} and \eqref{xi2}. 
Inserting this sum  in eq. \eqref{xi} gives the sought-after  contribution of  $O_{\rm reg}^{(4)}(M_{\rm o})$ to the 
renormalization of $O^{(2)}$. Adding the contribution from the cubic term, and collecting everything, 
leads to the  following expression for the renormalized  quadratic operator:
 \be\label{completeterm}
- \frac{2\pi^2}{k^2} {\rm tr} \left ( M_{\rm o}^c M_{\rm o}^d  +  \frac{i}{k}  f^{abc} [M_{\rm o}^a,  M_{\rm o}^b]  M_{\rm o}^d\,  \log \epsilon  +   \frac{2}{k} 
 M_{\rm o}^a [M_{\rm o}^a,  M_{\rm o}^c]  M_{\rm o}^d \,  \log \epsilon  + O \left (\frac{1}{k^2} \right)  \right )  J^c_0 J^d_0  \ .
\ee
The three  contributions in the above sum come, respectively, from  the quadratic, cubic and quartic terms  in the Taylor
expansion of  the defect-loop operator $O_{\rm reg}(M_{\rm o})$. 
We have used
 \be
\sum_{n > 0} \frac{e^{-|n|\epsilon}}{n} = - \log \epsilon + O(\epsilon) \; , 
\ee
and we have dropped terms which vanish in the limit $\epsilon\to 0$. 
 
\vskip 0.6mm

A simple calculation shows that the expression
\eqref{completeterm} can be put in the form
\eqref{QuadTermRenWL} if we define the renormalized couplings $M$ as follows:
\be
M^a = M_{\rm o}^a + \frac{1}{2k} \log\epsilon\,  \left [ M_{\rm o}^b,\,  i f^{abc} M_{\rm o}^c -  [M_{\rm o}^a, M_{\rm o}^b] \right ] + O \bigg ( \frac{1}{k^2} \bigg ) \;.
\ee
Taking the derivative with respect to $\log\epsilon$  gives the $\beta$-function of the running couplings. 
In principle,  to eliminate all $\epsilon$ dependence, the $\beta$-function should be expressed in terms of the renormalized 
rather than the bare couplings. To the leading order that concerns us here,  however, this is
a  trivial rewriting because $ M_{\rm o}^a  = M^a + O(1/k)$. We thus find:
 \be
\label{BetaFunction}
\beta^a(M) = - \frac{d M^a}{d \log\epsilon} = - \frac{1}{2k} \left [ M^b,\,   i f^{abc} M^c - [M^a, M^b]   \right ]  + O \bigg ( \frac{1}{k^2} \bigg ) \;.
\ee
These beta functions coincide with the gradient of the effective matrix action \eqref{EffOAct}, computed in 
ref.  \cite{Alekseev:2000fd}.  
We have essentially rederived this effective action from a closed-string point of view.
Note that the renormalization scale is implicitly the radius of the cylinder base which has been  set equal to 1,
so that the flow is in the sense of decreasing $\epsilon$. This explains the minus sign in the definition
of the $\beta$-function above.


\subsection{Fixed points and RG flows}
\label{fpRG}

From the expression \eqref{BetaFunction}  for the $\beta$  function it follows  immediately that the RG flow
preserves all global group symmetries.  This conclusion is actually valid at all orders of the $1/k$ expansion.
Indeed,  our regularization and renormalization scheme is such that 
  the $\beta$-function is given by   (i) products 
   of the matrices $M^a$, and (ii) contractions of their upper indices with either  the structure
   constants, or the Killing metric of  $\mathfrak{g}$.  All these operations are  covariant, so 
   any  global $H$ symmetry of the defect will be  preserved by the RG  flow,  as claimed.
    \vskip 0.5mm
    
    The situation is more subtle for  affine symmetries.  Recall from the analysis of
    section \ref{SecGlobRightSym} 
    that holomorphic  $H$-symmetric defects
    are parametrized by the invariant tensors $\Theta$ and $\tilde\Theta$. 
Classical 
$\widehat H$-invariance fixed $\Theta$  to be equal to the distinguished invariant tensor $\Theta_R$ associated with the representation $R$. 
 Using,  in the right-hand-side of  \eqref{BetaFunction},   the identities
\be
[ \Theta_{R}^j , \Theta_{R}^l ] = i f^{jli} \, \Theta_{R}^i\ \ \ {\rm and}\ \ [ \Theta_{R}^j  , \tilde\Theta^s] = if^{jsu} \tilde\Theta^u\ 
\ee
shows that   if  $\Theta  = \Theta_R$, then  $\beta^j=0$ for any $H$-invariant tensor $\tilde\Theta^s$.  Thus the
couplings  frozen by classical $\widehat H$ symmetry are not renormalized at this leading order, 
i.e. the subspace \eqref{fix} of parameter space
 is preserved by the RG flow.   
Furthermore, with  the help of the same identities,   one can  show  that  (for all $n\in \mathbb{Z}$ and $j= 1, \cdots , {\rm dim}\mathfrak{h}$) 
\be\label{widehatH}
   [ J^j_n, \, O_{\rm  ren}(\Theta_R, \tilde\Theta)] = 0 \ \   
\ee
at the first non-trivial order in the expansion (i.e. 
up to terms of order $1/k^3$).   At this order, the results of the classical analysis thus continue to hold.

\vskip 0.5mm
  
  How about higher orders in $1/k$? Since there are no chiral anomalies on the one-dimensional defect worldline, 
which would obstruct the gauge symmetry at the quantum level, 
 we expect that the affine $\widehat H$ symmetry  can still  be imposed. There is however, no reason to expect
that \eqref{widehatH} will continue to hold, and that $\beta^j =0$, 
 along  the classically-invariant subspace $\Theta  = \Theta_R$. 
The quantum--invariant subspace,  
$\Sigma(\widehat H, R, k)$, along which
$O_{\rm  ren}$ commutes with the algebra $\widehat {\mathfrak{h}}_{xk}$,  will  in general be  a  
deformation of  \eqref{fix}. It should be   defined by the vanishing of invariant tensors, $F$,  in $R\otimes R^* \otimes \mathfrak{h}$
which are built out of $\Theta$ and $\tilde\Theta$ and  which depend on the choice of renormalization scheme:
\be\label{forminvsub}
0 = F^{j} ( \Theta, \tilde\Theta) =  (\Theta^j - \Theta_R^j)  + O(1/k)\ \ .     
\ee
   Further support for  these claims  is provided
by the explicit construction  of the loop operator as a central element of a completion of
the enveloping algebra $U(\widehat {\mathfrak{h}}_{xk})$, at special points of the invariant subspace
which are fixed points of the RG flow  \cite{Alekseev:2007in}.   
This will be discussed in more detail in  section \ref{SecNonPert}.

\vskip 0.8mm

    Assuming that the above statements are valid, 
    we may   identify the WZW defects   in the invariant subspace \eqref{forminvsub}, and their RG flows, 
     with defects and flows in the $G/H$ coset model.  As a check of consistency, note that
  the defect-loop operators in the WZW model were by construction  invariant under rigid tanslations on the cylinder, i.e.     
             \be
     [  L_0^G \  ,\  O_{\rm ren}(M) ] \  = \  0\ \ \ \forall \ M\ . 
     \ee
 Using the GKO construction   \cite{Goddard:1984vk}   of the Virasoro
    algebra, 
    $L^{G/H}_n =  L_n^G - L_n^H$, and the fact that $L_n^G$ and $L_n^H$ are quadratic in the corresponding Kac-Moody currents, 
    one then  finds 
  \be
 [  L_{0}^{G/H}\  ,\  O_{\rm ren}(M) ]   \  = \   [  L_{0}^G\  ,\  O_{\rm ren}(M) ]  = 0
  \ee
  on the subspace $\Sigma(\widehat H, R, k)$
  on which  $ [ J^j_n, O_{\rm  ren}] = 0$. Our holomorphic loop operators are therefore also
  invariant under rigid translations in the  GKO model. Their RG flow  can  thus be
  imprinted smoothly on any boundary.
   
   \vskip 0.6mm
   
 Let us go back now to the flow equations \eqref{BetaFunction}. Its known fixed points, 
   analyzed in ref. \cite{Alekseev:2000fd, Alekseev:2002rj, Monnier:2005jt},  
  are essentially  of two types:  either the $M^a$ are commuting matrices,  or a subset of them 
    forms a representation of a subalgebra  $\mathfrak{h}\subseteq \mathfrak{g}$. The symmetry breaking fixed points with $\mathfrak{h} \neq \mathfrak{g}$
     were 
     described in \cite{Alekseev:2002rj, Monnier:2005jt}. 
    In particular, when  $R$ is the restriction of a representation of $G$,
      the following three fixed points are always
       present:\footnote{Other known fixed-point operators include (in an obvious notation)   $O_R^\mathfrak{h^\prime}$  for any
       proper subalgebra $ \mathfrak{h} \subset \mathfrak{h^\prime} \subset \mathfrak{g}$,  
       $O_r^\mathfrak{h}$ for any
         sub-representation $r\subset R$,  and marginal deformations thereof.}
   \be
     (0, 0)\  ,  \hskip 1cm  
   (T^j, 0)\ ,   \hskip 0.7cm  {\rm and}\ \ \ 
   (T^j, T^s )\ .   
   \ee
 where, as previously,  $M = (\Theta^j , \tilde\Theta^s)$ gives the  components of the
   adjoint vector (of matrices)  in  the linear subspaces $\mathfrak{h}$  and $\mathfrak{g}/\mathfrak{h}$,  
   and $T$
   are the normalized generators of $G$ in $R$.   We denote the corresponding loop operators respectively  by
   dim$(R)\, \mathbb{I}$ , $O_R^\mathfrak{h}$ and $O_R^\mathfrak{g}$. 
  It can  be proved  that the fixed point corresponding to $O_R^\mathfrak{g}$ 
  is stable when $R$ does not contain isomorphic $\mathfrak{g}$-subrepresentations \cite{Monnier:2005jt}, while the other two (in general) are not.  We also know, based on the analysis
  of \cite{Bachas:2004sy}, that    ${\rm dim}(R)\, \mathbb{I} \to O_R^\mathfrak{h}$ and 
  ${\rm dim}(R)\, \mathbb{I} \to O_R^\mathfrak{g}$ are generalized Kondo flows
  (where in the first one,  
   the defect couples only to the currents in the subalgebra $\mathfrak{h}$). We
  therefore expect the following flow diagram: 
\vskip 2mm
 
  \begin{equation}\label{flowchart}
\begin{diagram}
\node{ {\rm dim}(R)\, \mathbb{I} \ \ }
       \arrow[6]{e,t,..}{[H]}  \arrow[6]{se,r,..}{[G]}
    \node[8]{\ \ O_R^\mathfrak{h} } \arrow[5]{s,b}{[\widehat H]}\\[7]
\node[9]{O_R^\mathfrak{g}}
\end{diagram}
\end{equation}
 
\vskip 2mm
\noindent Indicated,  in square brackets,  are the symmetries that can be preserved along the flows. 
 The only  flow preserving the affine, $\widehat H$,  symmetry is indicated by a solid arrow. This class of flows
 descends from the WZW  to the $G/H$ model and provides, as we will see in section \ref{DerGenAfflLud},
 an explanation for the generalized Affleck-Ludwig rule \cite{Fredenhagen:2001kw, Fredenhagen:2002qn, Fredenhagen:2003xf}.

\vskip 0.5mm

 We will now exhibit the existence of these flows  in the two simple examples of section \ref{exfew}.  
 The study of the more general RG-flow  equation  \eqref{BetaFunction} presents no conceptual difficulties, and
 should be easy to  perform numerically.\footnote{A related first-order equation,  sharing the same non-abelian fixed points,
 is Nahm's generalized equation  $ dM^a/dt  =  M^a -  if^{abc} M^b M^c$,  which describes (among other things) 
  supersymmetric domain walls 
  \cite{Bachas:2000dx}. A topological classification is, in this case, possible by  analytic methods  \cite{Kronheimer:1990ay}. 
 }

\subsection{Two examples}
\label{Twoexamples}
 
 The first example has $G=SU(2)$, $H=O(2)$, and the defect transforming in the fundamental representation  of $SU(2)$. 
The most general Hamiltonian preserving the global $O(2)$ symmetry was given by eq. \eqref{ex2para1}, which we repeat
here for the readers' convenience:
$$
{\cal H}_{\rm imp}^{(1/2)}(\lambda,\tilde\lambda) \ = \  - \frac{1}{\sqrt{2}k}\,   \Bigl(
 \lambda \,  {\sigma^3}   {\cal J}_-^3     + \tilde\lambda\, ( {\sigma^1} \,      {\cal J}_-^1 
+   \,   {\sigma^2 } \,     {\cal J}_-^2)  \Bigr)\ . 
$$
Inserting this in the expression  \eqref{BetaFunction} for the $\beta$-function gives:
 \begin{align}\label{RGex1}
   {d\lambda\over dt} & \ = \ \frac{2}{k} \tilde\lambda^2(1 - \lambda)  + O\left({1\over k^2}\right) \nonumber \\ 
  {d\tilde\lambda\over dt}\  = \  &\frac{1}{k} \tilde\lambda \, ( 2 \lambda    - \lambda^2  - \tilde\lambda^2)  + O\left({1\over k^2}\right)
 \end{align}
 where $t = -{\rm log}\epsilon$ is the renormalization-group ``time".  The flow in this two-parameter space
 is shown in the left-hand side  of Figure 1.   Two 
   invariant subspaces are $ \lambda  = \tilde\lambda$, for global $SU(2)$ symmetry,  and  $\lambda = 1$ 
  for affine $\widehat O(2)$  symmetry.  The RG flows on these invariant subspaces (colored, respectively, in
  the figure in green and red) are the Kondo, and the Fredenhagen-Schomerus flows.  
  The full affine $\widehat{SU}(2)$ invariance is restored at their intersection. The RG equations \eqref{RGex1}  
  are also symmetric under the discrete transformations
   $\tilde \lambda \rightarrow -\tilde \lambda$, and  $\lambda \rightarrow 2 -\lambda$. The first
   is induced by the internal automorphism $\exp ( \pi i J^3_0/\sqrt{2} )$ of $\mathfrak{g}$. 
   The second is an accidental symmetry of the
action \eqref{EffOAct} for matrices restricted according to
the 2-parameter ansatz corresponding to the defect Hamiltonian \eqref{ex2para1}.
    
   Note also 
  that the  $\tilde\lambda=0$ axis  is a line of
  marginal deformations. This is not surprising, because at $\tilde{\lambda}=0$, the defect observable involves only $\sigma^3$ and no renormalization is necessary to make sense of the corresponding operator at the quantum level. Indeed, the path ordering has no effect and the integration of the Hamiltonian of the defect can be performed explicitly, yielding the zero mode of $J^3$ in the exponent. As a result the loop operator coincides with the multiplication by an element of the Cartan torus of $SU(2)$. This transformation is a global symmetry and is therefore marginal.   
 
\vskip 7mm 
 
\begin{figure}[!htb]
\centering
\includegraphics[scale=.84]{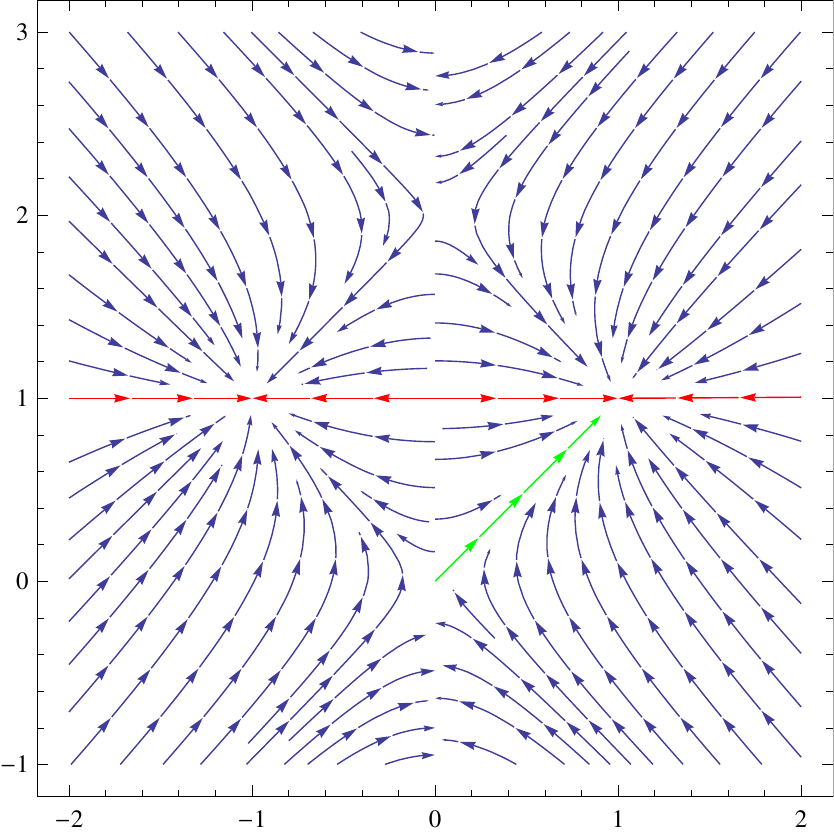} \hspace{1cm}
\includegraphics[scale=.8]{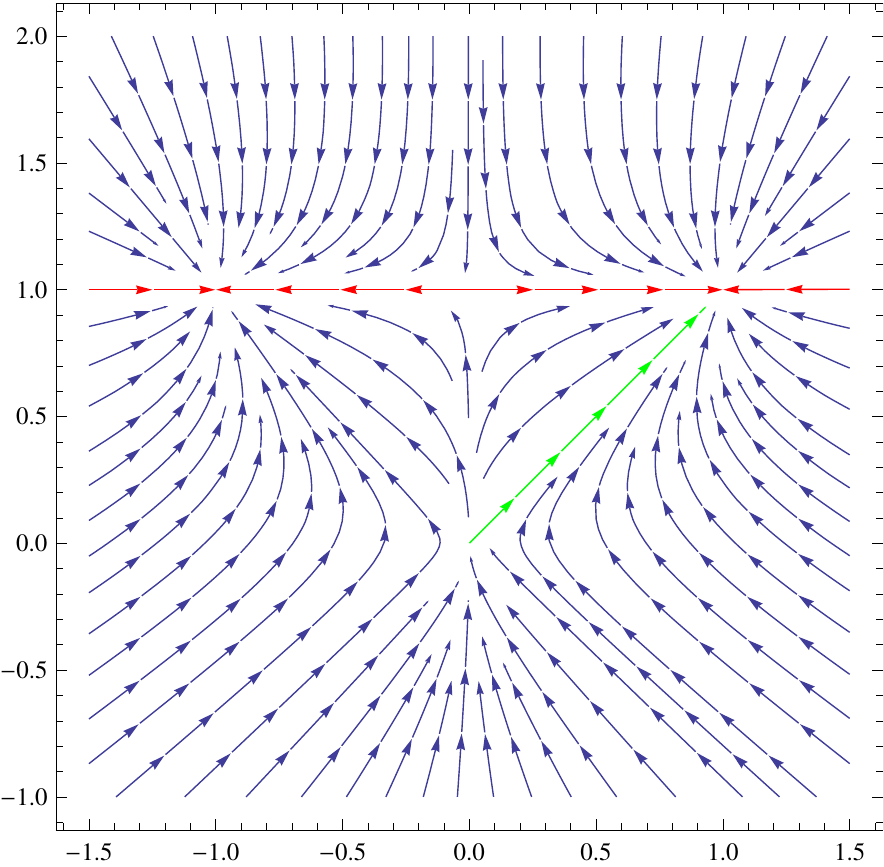}
\caption{\small  The pattern of flows for the two examples discussed in the text: 
 $G=SU(2)$, $H=O(2)$ ({\it left}) and $G=SU(2)\times SU(2)$, $H=SU(2)_{\rm diag}$ ({\it right}). 
$\lambda$  and $\tilde \lambda$ are plotted respectively along the vertical and horizontal axis. The Kondo flows of the $G$-symmetric defects ($\lambda = \tilde{\lambda}$),  analyzed in \cite{Bachas:2004sy}, 
are shown in green. The flows along the $\widehat H$-invariant subspaces ($\lambda = 1$) are drawn in red. They descend
to the Fredenhagen-Schomerus flows in the $G/H$ coset models.  
The symmetry $\tilde\lambda \rightarrow -\tilde\lambda$
 corresponds to automorphisms of the algebras $\mathfrak{g}$,  as explained in the text. 
 Both cases have fixed points at $(0,0), (0, 1)$ and $(1, \pm 1)$. 
 In the first example,  the first two fixed points lie on 
the  $\lambda$ axis, which is a line of marginal deformations. }
\label{Fig1}
\end{figure}

\eject
 

\vskip 0.8mm 
The second example of section  \ref{exfew} has  $G=SU(2)\times SU(2)^\prime$, $H=SU(2)_{\rm diag}$ and a defect
associated to a representation $R = (j,0)$ of $G$.  The general Hamitlonian \eqref{ex2para2} 
has  again  two free parameters. It reads  
$$
{\cal H}_{\rm imp}^{(j,0)}(\lambda,\tilde \lambda) = - \frac{1}{2k} \Bigl[
\lambda \,  \vec t \cdot  (\vec {\cal J}_-+ \vec {\cal J}_-^\prime) +
\tilde \lambda\,  \vec t \cdot  (\vec {\cal J}_- -  \vec {\cal J}_-^\prime)\Bigr] \ , 
$$
where $\vec t$ is the triplet of  $SU(2)$ generators in the representation of spin $j$. 
A straightforward calculation leads to  the following  $\beta$-function equations in this example:
 \begin{align}\label{84}
  {d\lambda_\pm \over dt} & \ = \ \frac{8}{k} \lambda_\pm (  \lambda_\pm - \lambda_\pm^2  - \lambda_\mp^2)  + O\left({1\over k^2}\right) \ ,  
   \end{align}
where $\lambda_\pm = (\lambda \pm \tilde\lambda)/2$ are the couplings to the two $SU(2)$ factors of $G$. 
The RG flow \eqref{84}    is exhibited in the right-hand-side of Figure 1. 
Local gauge invariance fixes $\lambda = \lambda_+ + \lambda_-=1$, leaving $\tilde\lambda$ as the only running coupling. 
On the diagonal  line,  $\lambda =  \tilde \lambda$,    the full  global $G$ symmetry is restored. 
Finally, on the $\lambda$ axis only the currents of $SU(2)_{\rm diag}$ enter in the defect  Hamiltonian. 
 The   flow  diagram \eqref{flowchart} is realized   on these three invariant lines.   The flow (in red) along  the line $\lambda = 1$ 
  is the only one that descends to the coset model. The symmetry under the exchange
   $ \lambda_+ \leftrightarrow    \lambda_-$
corresponds here to the outer automorphism exchanging the two factors of $G$.



\section{Exact quantization and boundary flows}
\label{SecNonPert}

The non-perturbative quantization of the holomorphic defects for generic $M^a$ 
  is at present an  open problem.  On the other hand, the special  fixed-point operators which correspond to the classical monodromies
 (see section \ref{LGI})  can  be quantized  along the lines 
  of   \cite{Alekseev:2007in} exactly. We will now explain how this works and then use the exact
  spectrum of the quantum monodromy traces to derive the 
   generalized Affleck-Ludwig rule \eqref{5}.


\subsection{Exact quantization of the fixed-point defects}
\label{exactfp}

Throughout this section, we choose $R$ to be an irreducible representation of $G$ of highest weight $\mu$, and write ${O}_\mu^\mathfrak{g}$ and ${O}_\mu^\mathfrak{h}$ for the two
 fixed-point loop operators denoted ${O}_R^\mathfrak{g}$ and ${O}_R^\mathfrak{h}$ in the previous section. Closely related loop operators
were constructed in ref. \cite{Alekseev:2007in} in the context of ungauged WZW models. We will apply here the same quantization technique to construct the relevant operators in the ungauged WZW model and then check that they have well-defined actions on the state space of the gauged model (i.e. are gauge invariant).

Let us consider first the stable, maximally-symmetric fixed point at 
$M^a = T^a$. The central idea in \cite{Alekseev:2007in} was to use two properties of the classical loop observable: it involves only the holomorphic current and it has vanishing Poisson bracket with it. Demanding these properties to be preserved by quantization, we ask that ${O}_\mu^\mathfrak{g}$ should be expressed as a series in the generators of $\hat{\mathfrak{g}}_k$ and that it commutes with all these generators. More precisely, ${O}_\mu^\mathfrak{g}$ should be a central element in an appropriate completion $\bar{U}(\hat{\mathfrak{g}}_k)$ of the enveloping algebra $U(\hat{\mathfrak{g}}_k)$. Building up on \cite{MR735060}, a recursive algorithm was presented in \cite{Alekseev:2007in} to compute the series expressing $O^{\mathfrak{g}}_\mu$ in terms of the currents $J^G$. As it is central, $O^{\mathfrak{g}}_\mu$ acts by scalar multiplication on each irreducible highest-weight module of $\hat{\mathfrak{g}}_k$. The quantization procedure also provides its eigenvalue:
\be
\label{SpecIR}
O^{\mathfrak{g}}_\mu = \chi^\mathfrak{g}_\mu \left ( -\frac{2\pi i}{k +\check{h}_\mathfrak{g}} (\nu + \rho_\mathfrak{g}) \right ) \mathbbm{1}
 \qquad \mbox{on } L^\mathfrak{g}_{(\nu, k)} \ .
\ee
Here $\chi^\mathfrak{g}_\mu$ is the character of $R$ seen as a function on the weight space of $\mathfrak{g}$, while
  $\check{h}_\mathfrak{g}$ and $\rho_\mathfrak{g}$ are  the dual Coxeter number and the Weyl vector
of $\mathfrak{g}$. In particular, if $\nu$ and $\mu$ are integrable weights of $\hat{\mathfrak{g}}_k$, then this expression simplifies to
\be
\label{SpecIRInt}
O^{\mathfrak{g}}_\mu =  \frac{S^\mathfrak{g}_{\mu\nu}}{S^\mathfrak{g}_{0\nu}} \mathbbm{1} \qquad \mbox{on } L^\mathfrak{g}_{(\nu, k)} \ ,
\ee
where  $S^\mathfrak{g}_{\mu\nu}$ is the modular $S$-matrix element of $\hat{\mathfrak{g}}_k$.
As noted in the introduction, and further discussed in subsection \ref{LGI},  
$O^{\mathfrak{g}}_\mu$ is the quantum analog of the trace of the classical monodromy matrix (for early semi-classical studies, see  \cite{Gepner:1986wi, Alekseev:1990vr}).
 
\vskip 0.5mm

A comment is in order here concerning the two fixed points, at  $\tilde\lambda = \pm 1$, 
 in the $SU(2)/O(2)$ example of section \ref{Twoexamples}. 
Recall that the map  $\tilde\lambda \rightarrow -\tilde\lambda$ can be implemented by conjugating the $SU(2)$
currents  with  the operator $\exp \left(\frac{\pi i}{\sqrt{2}}J^3_0 \right)$. Since this commutes with  
 $O^{\mathfrak{g}}_\mu$, we conclude that the loop operators at these two fixed points coincide. 

\vskip 0.5mm

Let us come now to the unstable fixed point at $(\Theta^j, \tilde\Theta^s) =  (T^j, 0)$. Again, the corresponding classical loop observable is characterized by two facts: it is built out of the restriction of the WZW current to the subgroup $H$ and its Poisson bracket with the restricted current vanishes. We deduce that $O^{\mathfrak{h}}_\mu$ has to be an element of the center of the completed enveloping algebra $\bar{U}(\hat{\mathfrak{h}}_{kx})$ and it can be constructed along the same lines. Such operators were considered in \cite{Alekseev:2007in} as loop operators breaking the chiral symmetry of the WZW model from $\hat{\mathfrak{g}}$ to $\hat{\mathfrak{h}} \times \hat{\mathfrak{g}}/\hat{\mathfrak{h}}$, where the second factor denotes the coset vertex algebra. The action of the quantum operator, $O^{\mathfrak{h}}_\mu$, 
on a highest weight $\hat{\mathfrak{h}}$-module $L^\mathfrak{h}_{(\beta, \kappa)}$   reads
 \be
\label{SpecUV}
 O^{\mathfrak{h}}_\mu =  \sum_\alpha   b_{\mu\alpha}\,  O^{\mathfrak{h}}_\alpha = \sum_\alpha   b_{\mu\alpha}\, 
   \chi^\mathfrak{h}_\alpha \left ( -\frac{2\pi i}{\kappa+\check{h}_\mathfrak{h}}
 (\gamma + \rho_\mathfrak{h}) \right ) \mathbbm{1} \qquad \mbox{on } L^\mathfrak{h}_{(\gamma, \kappa)}\ ,  
\ee
where  $\kappa$ is the level and $\gamma$ is an integrable highest weight of 
the Kac-Moody algebra $\hat {\mathfrak{h}}_{\kappa}$. Furthermore,  
$b_{\mu\alpha}$ are the branching coefficients in the decomposition of 
$R$ in irreducible  $\mathfrak{h}$-representations of highest weight $\alpha$. Note that here and in what follows we use Greek letters from the middle of
the alphabet to denote highest weights, and corresponding representations of $\mathfrak{g}$, while  letters from
the beginning of the alphabet  stand for highest weights and representations of  the subalgebra $\mathfrak{h}$.

\vskip 0.5mm

Let us now check the gauge invariance of the operators constructed. The coset modules, which form the building blocks of the state space of the coset model, are constructed by taking the quotient by the action of the Kac-Moody subalgebra $\hat{\mathfrak{h}} \subseteq \hat{\mathfrak{g}}$. Any operator commuting with the action of $\hat{\mathfrak{h}}$ has a well-defined action on the coset modules, and therefore on the state space of the coset model. But $O^{\mathfrak{g}}_R$ and $O^{\mathfrak{h}}_R$ commute with $\hat{\mathfrak{h}} \subseteq \hat{\mathfrak{g}}$ by construction. As a matter of fact, $O^{\mathfrak{g}}_R$ commutes with $\hat{\mathfrak{g}}$ and $O^{\mathfrak{h}}_R$, being a central element in $\hat{U}(\hat{\mathfrak{h}}_{kx}) \subseteq \hat{U}(\hat{\mathfrak{g}}_{k})$, commutes with the action of $\hat{\mathfrak{h}}$. Using the decomposition \eqref{DecompGMod} of the integrable modules of $\hat{\mathfrak{g}}_k$ in terms of modules of $\hat{\mathfrak{h}}_{xk}$ and modules of the coset vertex algebra, we deduce that the actions of the two fixed point operators are given by
 \be
 \label{92}
O^{\mathfrak{g}}_\mu  = \frac{S^\mathfrak{g}_{\mu\nu}}{S^\mathfrak{g}_{0\nu}} \mathbbm{1} \ , \qquad \mbox{and} \ \ \ 
O^{\mathfrak{h}}_\mu  =  \sum_\alpha b_{\mu\alpha} \,  \frac{S^\mathfrak{h}_{\alpha\gamma}}{S^\mathfrak{h}_{0\gamma}} \mathbbm{1} \qquad
 \mbox{on }\ \  L^{\mathfrak{g}/\mathfrak{h}}_{[\nu,\gamma]} \ , 
\ee
 where $S^\mathfrak{g}$ and $S^\mathfrak{h}$ are the modular $S$-matrices for $\mathfrak{g}_{k}$ and $\mathfrak{h}_{xk}$, and $b_{\mu\alpha}$ the branching
 coefficients. Remark that they both act by scalar multiplication on the coset modules, and therefore commute with the action of the coset vertex algebra.
\vskip 0.5mm

In appendix \ref{BRST} we will show how to rederive the above results in the BRST quantization of the gauged WZW model.


 \subsection{The generalized Affleck-Ludwig rule}
\label{DerGenAfflLud}

We have argued in this paper that there exist RG flows of holomorphic defects in the GKO model based on the coset $G/H$, 
which send 
\be
\label{RGFowLoopOp2}
O^{\mathfrak{h}}_\mu  \mapsto  O^{\mathfrak{g}}_\mu \ ,
\ee
for any $G$-representation of highest weight $\mu$. 
These are the flows marked in red in the examples of figure 1. We have computed explicitly these flows at the leading order in $1/k$, and the
 loop operators at the two fixed points exactly.  We have also explained in section \ref{SecPertQuant} how the renormalized operator along
the flow can be made to commute with translations on the cylinder. Following the ideas of \cite{Graham:2003nc, Bachas:2004sy},
we can thus push the defect to a conformal boundary and induce automatically a boundary RG flow.  We will
now explain how the flow \eqref{RGFowLoopOp2} can be used to derive the generalized Affleck-Ludwig rule
 \cite{Fredenhagen:2001kw, Fredenhagen:2002qn, Fredenhagen:2003xf}, i.e. expression
\eqref{5} of the introduction. 

\vskip 0.5mm 

We start by reviewing the construction of boundary states in coset models (see \cite{Fredenhagen:2003xf} for more details). 
Recall that the primary fields  are labeled by pairs $(\mu,\alpha)$ of integrable weights for $\hat{\mathfrak{g}}_k$  and $\hat{\mathfrak{h}}_{xk}$.
 Not all of these pairs are admissible, and some pairs must be identified (see for instance \cite{CFT1997}, chapter 18). Let us write $[\mu,\alpha]$ for the equivalence 
 classes of the admissible pairs, and denote by $\mathcal{P}$ the set of these equivalence classes.
 The modular S-matrix of the coset theory can be expressed in term of the modular S-matrices  of $\hat{\mathfrak{g}}_k$ and $\hat{\mathfrak{h}}_{xk}$ as follows :
\be
\label{SMatCoset}
S^{\mathfrak{g}/\mathfrak{h}}_{[\mu,\alpha],[\nu,\beta]} = n S^\mathfrak{g}_{\mu\nu} (S^\mathfrak{h})^{-1}_{\alpha\beta} \ ,
\ee
where we chose particular representatives $(\mu,\alpha)$ and $(\nu,\beta)$ in the equivalence classes $[\mu,\alpha]$ and $[\nu,\beta]$, 
and $n$ denotes the number of pairs in each equivalence class.\footnote{To simplify the formulae, we
 are assuming that this number is constant. This is the case in many important examples of coset models.}
The right-hand side of \eqref{SMatCoset} is, of course,  independent of this choice.
 
 \vskip 0.5mm 
 
We can construct conformal boundary states for this theory, starting with any representation of the fusion algebra
by  matrices with non-negative integer entries. 
 More specifically, we consider a
 set of matrices $N_{[\mu,\alpha]}$,  indexed by the primary fields of our coset model, 
 with non-negative integer matrix elements 
 $(N_{[\mu,\alpha]})_I^{\  J}$. These must  form a representation of the fusion algebra:
\be
N_{[\mu,\alpha]} \  N_{[\nu,\beta]} = \sum_{[\rho, \gamma]} \mathcal{N}_{[\mu,\alpha],[\nu,\beta]}^{\qquad [\rho, \gamma]} \  N_{[\rho, \gamma]} \ ,
\ee
where $\mathcal{N_{\rm XY}^{\ \ \rm Z}}$ are the fusion coefficients. They must furthermore obey the conditions
\be
N_{[0,0]} = \mathbbm{1} \qquad \mbox{and }\ \ \ N_{[\mu,\alpha]} = (N_{[\mu^\ast,\alpha^\ast]})^T \ ,
\ee
where $\mu^\ast$ is the weight conjugate to $\mu$. 
 Now since the fusion rules form a commutative algebra, 
all of its irreducible representations are one-dimensional. The latter are labeled by the coset fields $[\nu,\beta]$ and given explicitly by 
\be
[\mu,\alpha] \mapsto \frac{S^{\mathfrak{g}/\mathfrak{h}}_{[\mu,\alpha],[\nu,\beta]}}{S^{\mathfrak{g}/\mathfrak{h}}_{[0,0],[\nu,\beta]}} \ .
\ee
The representation provided by $N$ decomposes into these irreducible representations as follows:
\be
\label{DecompNimRepIrred}
\sum_J (N_{[\mu,\alpha]})_I^{\ J} \ \psi_J^{[\nu,\beta],i} =
 \frac{S^{\mathfrak{g}/\mathfrak{h}}_{[\mu,\alpha],[\nu,\beta]}}{S^{\mathfrak{g}/\mathfrak{h}}_{[0,0],[\nu,\beta]}}\  \psi_I^{[\nu,\beta],i} \ ,
\ee
where $\psi^{[\nu,\beta],i}$ are the common eigenvectors of the matrices $N$. We included, for completeness, 
 $i$ as a possible multiplicity index,  and we denote this multiplicity by $n_{[\nu,\beta]}$. {\it Attention}: the indices in this 
 subsection should not
 be confused with those in the rest of the paper.  
 \vskip 0.8mm

Consider next the torus partition function of the  coset model  
$$
Z(q,\bar{q}) = Z_{[\mu,\alpha],[\nu,\beta]} \, \chi^{\mathfrak{g}/\mathfrak{h}}_{[\mu,\alpha]}(q)\,  \chi^{\mathfrak{g}/\mathfrak{h}}_{[\nu,\beta]}(\bar{q}) \ ,
$$ 
where $\chi^{\mathfrak{g}/\mathfrak{h}}_{[\mu,\alpha]}(q)$ is the character of the coset module $L^{\mathfrak{g}/\mathfrak{h}}_{[\mu,\alpha]}$. 
 Suppose moreover that there exists an automorphism $\Omega$ of the coset vertex algebra,
  such that $Z_{[\nu,\beta],\Omega[\nu,\beta]} = n_{[\nu,\beta]}$.    
   We can then build an elementary boundary state for each value of the matrix index  $I$ :
\be
\label{CosetBoundState}
 B_I  \ \equiv\  \sum_{[\nu,\beta] \in \mathcal{P}}\  \frac{\psi^{[\nu,\beta],i}_I}{\sqrt{S^{\mathfrak{g}/\mathfrak{h}}_{[0,0],[\nu,\beta]}}} \ket{[\nu,\beta],i} \! \rangle \ ,
\ee
where $\ket{[\nu,\beta],i} \! \rangle$ are the Ishibashi states corresponding to the $n_{[\nu,\beta]}$ orthogonal coset modules, 
 isomorphic to $L^{\mathfrak{g}/\mathfrak{h}}_{[\nu,\beta]} \otimes L^{\mathfrak{g}/\mathfrak{h}}_{\Omega[\nu,\beta]}$,    in the closed-string spectrum.

\vskip 0.8mm

We are at long last ready to compute the RG flows induced on conformal boundaries by our holomorphic-defect
flow \eqref{RGFowLoopOp2}. Acting with the UV and IR  fixed-point operators on the conformal boundary state $B_I$, and inserting the
exact formulae  \eqref{92},   leads to: 
 \be
\sum_\alpha b_{\mu\alpha} \sum_{[\nu,\gamma] \in \mathcal{P}} \frac{S^\mathfrak{h}_{\alpha\gamma}}{S^\mathfrak{h}_{0\gamma}} \frac{\psi^{[\nu,\gamma],i}_I}{\sqrt{S^{\mathfrak{g}/\mathfrak{h}}_{[0,0],[\nu,\gamma]}}} \ket{[\nu,\gamma],i} \! \rangle 
\mapsto 
\sum_{[\nu,\gamma] \in \mathcal{P}}   
\frac{S_{\mu\nu}}{S_{0\nu}}
\frac{\psi^{[\nu,\gamma],i}_I}{\sqrt{S^{\mathfrak{g}/\mathfrak{h}}_{[0,0],[\nu,\gamma]}}} \ket{[\nu,\gamma],i} \! \rangle \ .
\ee
Using the identities (which follow from \eqref{SMatCoset})
\be
\left( \frac{S^\mathfrak{h}_{\alpha\gamma}}{S^\mathfrak{h}_{0\gamma}}\right)^{-1}\
 =\  \frac{S^{\mathfrak{g}/\mathfrak{h}}_{[0,\alpha],[\nu,\gamma]}}{S^{\mathfrak{g}/\mathfrak{h}}_{[0,0],[\nu,\gamma]}} \qquad \mbox{and} \qquad 
\frac{S^\mathfrak{g}_{\mu\nu}}{S^\mathfrak{g}_{0\nu}} = \frac{S^{\mathfrak{g}/\mathfrak{h}}_{[\mu,0],[\nu,\gamma]}}{S^{\mathfrak{g}/\mathfrak{h}}_{[0,0],[\nu,\gamma]}} \ ,
\ee
as well as the definition \eqref{DecompNimRepIrred} of the coefficients $\psi$, we finally find the boundary flows:
\be
\label{FlowIndFromLoopOp}
\sum_\alpha b_{\mu\alpha} \sum_J (N_{[0,\alpha]})_I^{\ J}  B_J \ \ \mapsto\ \  \sum_J (N_{[\mu,0]})_I^{\ J}  B_J  \ .
\ee
This is precisely  the generalization of the Affleck-Ludwig rule to coset models described in refs. \cite{Fredenhagen:2001kw, Fredenhagen:2002qn, Fredenhagen:2003xf}. 
The RG flows  \eqref{5}  of the introduction are a special instance of the above rule, where  $N_{[\mu,0]}$  are replaced by the fusion matrices
${\cal N}_{[\mu,0]}$
 (and the matrix indices $I,J \cdots$ run over  $\mathcal{P}$).
 The fusion matrices  form themselves a representation of the fusion
algebra,  and the boundary states constructed out of them are the maximally symmetric boundary states first constructed by Cardy \cite{Cardy:1989ir}. Note, as a consistency check, that the coefficients of the boundary states on both sides of \eqref{FlowIndFromLoopOp} are non-negative integers.
 For a detailed account of the boundary RG flows falling in the scope of the generalized Affleck-Ludwig rule in the critical Ising model,
 the tricritical Ising model and the three states Potts model, see the appendix B of \cite{Fredenhagen:2003xf}. 
Our derivation of this rule from  flows of holomorphic bulk defects shows its  universality, in particular the fact that the beta functions are independent of the boundary condition. 
\vskip 0.5mm

 Let us note in closing that bulk defects of  a  conformal field theory,  A, can be viewed  via  the folding trick
\cite{Oshikawa:1996dj, Bachas:2001vj} as boundaries
of  the theory  A$\otimes {P}({\rm A})$,  where ${P}$ is the parity transformation.  
Equations \eqref{fusion} on the other hand suggest that,  in certain situations,  holomorphic defects can be 
in one-to-one correspondence with boundaries of the ({\it single-copy}) theory A. ``Under what conditions does this happen?"
is a question that deserves further investigation.
 

\vskip 0.4cm
{\bf Aknowledgements}
\vskip 2mm
We thank Anton Alekseev, Ilka Brunner, Denis Bernard, Ingo Runkel, Samson Shatashvili and Kostya  Zarembo for discussions. 
This work has been supported by the European Networks 'Superstring Theory' (MRTN-CT-2004-512194) and `Forces Universe' (MRTN-CT-2004-005104)
and by the Agence Nationale pour la Recherche (contract  05-BLAN-0079-01). 
S.M. is supported in part by the fellowship PBGE2--121187 of the Swiss National Science Foundation.


\appendix
\section{Quantum monodromies in the BRST scheme}
\label{BRST} 
 
We have seen in this paper that the endpoints of the
Fredenhagen-Schomerus flows  were defects with  classical couplings  $(\Theta^j , \tilde \Theta^s) = (T^j, 0)$
and $(T^j, T^s)$.  In section \ref{exactfp} we have constructed, in the GKO quantization scheme, 
 the corresponding quantum-monodromy operators
$O^{\mathfrak{h}}_\mu$ and  $O^{\mathfrak{g}}_\mu$. We will  here show that the same
result can be obtained in the BRST quantization. The starting point in the BRST scheme is the
 expression \eqref{flatcc} for the impurity Hamiltonian, with the choice $\tilde \Theta^j = T^j$
 for the unphysical couplings.  With this choice, at the two endpoints of the flow the defect couples only to the
 currents, respectively,  of $H$ and $G$. The two fixed-point operators are thus central elements
 in the enveloping algebras of the two independent Kac-Moody algebras:  $\hat{\mathfrak{g}}$ at level $k$
 and $\hat{\mathfrak{h}}$
at level $-xk - 2\check{h}_\mathfrak{h}$. We will show that these fixed-point operators have the same
action as their counterparts in the GKO quantization of the model.
 \vskip 0.4mm
 
Let us first review briefly how one derives in the BRST scheme the GKO space of states described 
    in section  \ref{statespace}.
For more details we refer the reader to   \cite{1990NuPhB.329..649K, Hwang:1993nc}.
   Recall  that we have an embedding $\hat{\mathfrak{h}} \subseteq \hat{\mathfrak{g}}$ of Kac-Moody algebras for
 each embedding of finite semisimple Lie algebras $\mathfrak{h} \subseteq \mathfrak{g}$. 
Accordingly,  a highest weight integrable module $L^\mathfrak{g}_{(\nu, k)}$ for $\hat{\mathfrak{g}}$ at level $k$ 
decomposes into $\hat{\mathfrak{h}}$-modules at level $xk$ according to eq. \eqref{DecompGMod}, 
$$
L^\mathfrak{g}_{(\nu, k)} = \bigoplus_{\gamma} L^\mathfrak{h}_{(\gamma, xk)} \otimes L^{\mathfrak{g}/\mathfrak{h}}_{[\nu,\gamma]} \ .  
$$
  The coset modules $L^{\mathfrak{g}/\mathfrak{h}}_{[\nu,\gamma]}$   carry an action of the coset vertex algebra, 
  which is composed of all the normal-ordered products of generators of $\hat{\mathfrak{g}}$  that commute with every element in $\hat{\mathfrak{h}}$.
   These modules are the  building blocks of the state space of the GKO coset models.

 \vskip 0.4mm
 
 By contrast  in the BRST scheme  the
  classical currents $\mathcal{J}^G$ and $\mathcal{J}^H$ are
  quantized separately into generators $J^G$ and $J^H$, and  
  the gauged WZW model starts out as a  direct  product of three  non-interacting theories:
 \begin{itemize}
	\item a WZW theory based on $G$ at level $k$, with current $J^G$; 
	\item  a 
WZW theory based on $H$ at level $-xk - 2\check{h}_\mathfrak{h}$, with current $J^H$, where  $x$ is the 
embedding index of $\mathfrak{h}$ in $\mathfrak{g}$,  and 
 $\check{h}_\mathfrak{h}$ is the dual Coxeter number of $\mathfrak{h}$; and
 \item
 a system of ghosts in the adjoint representation of $\mathfrak{h}$.
 \end{itemize}
 As a result, the extended space  before the  BRST projection to physical states,  is made out  of
  tensor products of three highest-weight modules:  
 \be
\label{ChirModBefBRST}
  L^\mathfrak{g}_{(\nu, k)} \otimes L^\mathfrak{h}_{(-\gamma-2\rho_\mathfrak{h}, -xk-2\check{h}_\mathfrak{h})} \otimes L^{\mbox{\tiny ghosts}} \ .
\ee
Here $\nu$ is an integrable weight of $\hat{\mathfrak{g}}_k$,  $\gamma$ an integrable weight of $\hat{\mathfrak{h}}_{xk}$,  
$\rho_{\mathfrak{h}}$ is the  Weyl vector  of  ${\mathfrak{h}}$ (the half sum of its positive roots), 
 and  
  $L^{\mbox{\tiny ghosts}}$ is a module for the ghost algebra which  plays a spectator role in  our discussion.
 It is understood  that ${J}^G$ acts on the first factor and ${J}^H$ on the second one.
  A basis of representatives of the relative BRST cohomology of this module is provided by states such that \cite{Hwang:1993nc}:
\begin{itemize}
	\item their component in the first factor lies in a given summand 
	$L^\mathfrak{h}_{(\gamma, xk)} \otimes L^{\mathfrak{g}/\mathfrak{h}}_{[\nu,\gamma]}$ of \eqref{DecompGMod}, and is of the form $\ket{0,\gamma,xk} \otimes \ket{\phi}$, 
	where the first state is the
	 highest weight vector of $L^{\mathfrak{h}}_{(\gamma,xk)}$ while  the second  is an arbitrary state of $L^{\mathfrak{g}/\mathfrak{h}}_{[\nu,\gamma]}$\ ; 
	\item their component in the second factor is the highest weight vector 
	$ \ket{0, -\gamma - 2 \rho_\mathfrak{h}, -xk -2 \check{h}_\mathfrak{h}}$;   
		\item their ghost component is the ghost vacuum $\ket{0, \mbox{ghosts}}$. 
\end{itemize}
It follows rather easily  that the relative BRST cohomology of \eqref{ChirModBefBRST} is isomorphic to $L^{\mathfrak{g}/\mathfrak{h}}_{[\nu,\gamma]}$. 
As the latter is the building block of the GKO construction of the state space, 
the equivalence of the BRST and  GKO approaches at the level of state space can be established.

\vskip 1.6mm

Let us now go back to the fixed-point operators $O^{\mathfrak{g}}_\mu$ and $O^{\mathfrak{h}}_\mu$. 
To each state $\ket{\phi} \in L^{\mathfrak{g}/\mathfrak{h}}_{[\nu,\gamma]}$ in the GKO construction corresponds a representative
\be
\label{TypStateGaugWZW}
\ket{\phi_{\rm BRST}} = \ket{0, \gamma, xk} \otimes \ket{\phi} \otimes \ket{0, -\gamma - 2 \rho_\mathfrak{h}, -xk -2 \check{h}_\mathfrak{h}} \otimes \ket{0, \mbox{ghosts}}
\ee
in the relative BRST cohomology of the gauged WZW model.
The IR operator  $O^{\mathfrak{g}}_\mu$ acts on the first two factors of  \eqref{TypStateGaugWZW}, i.e. on a state
in the  module   $L^\mathfrak{g}_{(\nu, k)}$ of $\mathfrak{g}_k$. Its action is thus given by \eqref{SpecIR}, i.e. it is the same
as in the GKO quantization.

 \vskip 0.9mm
 
The story is more interesting for the UV operator $O^{\mathfrak{h}}_\mu$ which acts on the third factor in the decomposition \eqref{TypStateGaugWZW}. 
Our  first  task is to establish  that the central elements constructed in  \cite{MR735060, Alekseev:2007in} 
have a well-defined action on the Kac-Moody modules for $\hat{\mathfrak{h}}$ at the subcritical level $-\check{h}_{\mathfrak{h}}-2k$.
 Let us recall how  central operators in a completion of the universal enveloping algebra of $\hat{\mathfrak{h}}$ were constructed by a recursive procedure
in these references.  First consider the triangular decomposition of $\hat{\mathfrak{h}}$ : $\hat{\mathfrak{h}} = \hat{\mathfrak{h}}_+ \oplus \hat{\mathfrak{h}}_0 \oplus \hat{\mathfrak{h}}_-$, where $\hat{\mathfrak{h}}_0$ is the Cartan subalgebra. Define the following functions on the affine weight space $(\hat{\mathfrak{h}}_0)^\ast$ :
 $$
T_{\omega}(\alpha) = 2(\alpha + \hat{\rho}_{\mathfrak{h}} - \omega, \omega) \ ,
$$
where the index $\omega$ is an element of the affine root lattice, and $\hat{\rho}_{\mathfrak{h}}$ is the affine Weyl vector of $\hat{\mathfrak{h}}$. (Recall that the affine Weyl vector has a horizontal part equal to the Weyl vector $\rho_{\mathfrak{h}}$ of the horizontal Lie algebra $\mathfrak{h}$ and a level equal to the dual Coxeter number $\check{h}_{\mathfrak{h}}$.) Consider also the subset $L \subset (\hat{\mathfrak{h}}_0)^\ast$, defined by $\alpha \in L \Leftrightarrow T_{n\psi}(\alpha+\varphi) \neq 0$ for each positive affine root 
 $\psi$, each positive integer $n$, and $\varphi$ in the affine root lattice. 
 Given any function $f$ analytic on $L$, it is possible to construct a central operator which has a well defined action on every module with highest weight $\alpha \in L$. The central operator acts on the module with highest weight $\alpha$ by scalar multiplication by $f(\alpha)$. Note that these modules are exactly the irreducible Verma modules of the Kac-Moody algebra, and that $L$ contains neither the integrables weights nor the weights at negative level appearing in the state space of the gauged WZW model.

In order to obtain a central operator that is well-defined on integrable modules, 
additional conditions must be imposed on the function $f$ defining the operator. These conditions take the following form \cite{MR735060} : 
\be
\label{NecCondExtCentOp}
T_{n\psi}(\alpha)=0 \qquad \Rightarrow \qquad f(\alpha) = f(\alpha - n\psi)
\ee
for $\psi$ a real positive root.
A slightly stronger but more conceptual condition is that $f$ should be invariant under the action of the Weyl group shifted by the Weyl vector $\rho_\mathfrak{h}$. One can easily check that the function \eqref{SpecIR} used to define Wilson operators satisfies this property. When the conditions \eqref{NecCondExtCentOp} are satisfied, it is proved in \cite{MR735060} (Theorem 2) that the central operator has a well defined action on any highest weight module whose highest weight lies in $-\hat{\rho}_\mathfrak{h}+K$, where $K$ is the Tits cone of $\hat{\mathfrak{h}}$. The Tits cone is defined as the set of all the weights which have negative scalar product with a finite number of roots only. The set $-\hat{\rho}_\mathfrak{h}+K$ does contain all the integrable weights of $\hat{\mathfrak{h}}$, but not our subcritical level weights. Indeed, a weight with negative level never belongs to $K$, because its scalar product with roots of sufficiently high grade would always be negative. Therefore, weights with level below the critical level $-\check{h}_\mathfrak{h}$ cannot belong to $-\hat{\rho}_\mathfrak{h}+K$, and we cannot use Theorem 2 of \cite{MR735060} to conclude that our operators have a well-defined action on the $\hat{\mathfrak{h}}$-modules  entering the BRST quantization of the gauged WZW model.

However, a careful analysis of the proof of this theorem shows that whenever the condition \eqref{NecCondExtCentOp} is satisfied, the action of central operators can be extended across the planes $T_{n\psi}(\alpha) = 0$ for any $\psi$ such that $(\psi,\psi) > 0$. So it may be impossible to perform this extension only on weights satisfying $T_{n\Psi}(\alpha) = 0$, where $\Psi$ is the imaginary simple root of the Kac-Moody algebra.  The scalar product of $\Psi$ with an affine weight gives the level of this weight, so for a generic affine weight $\alpha$ of level $\kappa$, we have 
\be
T_{n\Psi}(\alpha) = 2(\alpha+\hat{\rho}_{\mathfrak{h}} - n\Psi ,n\Psi) = 2n(\kappa + \check{h}_{\mathfrak{h}}) \ .
\ee
The extension may therefore fail only for weights with critical level $\kappa = -\check{h}_{\mathfrak{h}}$. This is not the case 
in the problem at hand, since $k > 0$ so that 
 our weights  are strictly subcritical.  We conclude 
 that the quantum-monodromy operators do have a well-defined action on each of the modules appearing in the state space of the gauged WZW model
 in the BRST quantization.
 \vskip 0.5mm   
 
 This rather lengthy argument shows that we can compute the action of 
  $O^{\mathfrak{h}}_\mu$ on the third factor of  \eqref{TypStateGaugWZW} by using
  eq.  \eqref{SpecUV} with the replacements:  $\gamma \to  -\gamma - 2\rho_\mathfrak{h}$ and  
   $\kappa \to  -xk -2 \check{h}_\mathfrak{h}$. 
 From the trivial  identity
 \be
 -\frac{2\pi i}{xk+\check{h}_\mathfrak{h}} (\gamma + \rho_\mathfrak{h})\,  =\,   -\frac{2\pi i}{(-xk -2 \check{h}_\mathfrak{h})
 +\check{h}_\mathfrak{h}} \left((-\gamma - 2\rho_\mathfrak{h}) + \rho_\mathfrak{h}\right)
 \ee
 we then conclude that the action of this fixed-point operator   is the same in the GKO and BRST quantizations of the
 gauged WZW model.


{\small
\providecommand{\href}[2]{#2}\begingroup\raggedright\endgroup
}
\end{document}